\documentclass[letterpaper,twocolumn,10pt]{article}
\usepackage{usenix}

\usepackage{xspace}

\usepackage{booktabs}

\usepackage{graphicx}
\usepackage{amsmath}

\usepackage[capitalize]{cleveref}

\usepackage[shortlabels,inline]{enumitem}

\usepackage{algorithm}
\usepackage[noend]{algpseudocode}

\usepackage{tikz}
\usetikzlibrary{shapes.geometric,positioning,arrows.meta,fit,calc,backgrounds}

\Crefname{ALG@line}{Line}{Lines}

\makeatletter
\algnewcommand{\LineComment}[1]{\Statex\hskip\ALG@thistlm{\color{gray}\textrm{// #1}}}
\makeatother

\def\sysname{Ira\xspace}

\newif\iffullversion
\fullversiontrue

\newcommand{\parhead}[1]{\vspace{0.5em}\noindent\textbf{#1.}\xspace}
\newcommand{\italhead}[1]{\vspace{0.2em}\noindent\textit{#1.}\xspace}

\usepackage[available]{usenixbadges}
\newcommand{\SloadCount}{696,683,837}  
\newcommand{\SstoreCount}{205,091,770}  
\newcommand{\StorageOpsTotal}{901,775,607}  
\newcommand{\AccountMetadataTotal}{6,751,623}  
\newcommand{\CallsTotal}{281,933,575}  
\newcommand{\CreationDestructionTotal}{2,427,834}  
\newcommand{\OpsGrandTotal}{1,192,888,639}  
\newcommand{\StorageOpsPct}{75.6}  
\newcommand{\AccountMetadataPct}{0.6}  
\newcommand{\CallsPct}{23.6}  
\newcommand{\CreationDestructionPct}{0.2}  
\newcommand{\StorageRwRatio}{3.4}  

\newcommand{\IoTraceBlockCount}{100,800}  
\newcommand{\IoTimePct}{67.9}  
\newcommand{\ComputeTimePct}{32.1}  
\newcommand{\IoComputeRatio}{2.1}  
\newcommand{\MedianIoTimePct}{66.8}  
\newcommand{\IoTimePctPFive}{55.6}  
\newcommand{\IoTimePctPNinetyFive}{80.2}  

\newcommand{\IntraBlockTotalKeys}{200,839,090}  
\newcommand{\IntraBlockTotalAccesses}{901,876,518}  
\newcommand{\IntraBlockSingleAccessPct}{34.5}  
\newcommand{\IntraBlockMultiAccessPct}{65.5}  
\newcommand{\IntraBlockRepeatedAccessPct}{77.7}  
\newcommand{\IntraBlockMaxAccessCount}{23,406}  
\newcommand{\IntraBlockKeysOne}{69,264,600}  
\newcommand{\IntraBlockKeysTwo}{63,445,255}  
\newcommand{\IntraBlockKeysThreeToFive}{49,614,494}  
\newcommand{\IntraBlockKeysSixToTen}{12,665,780}  
\newcommand{\IntraBlockKeysGtTen}{5,848,961}  
\newcommand{\IntraBlockKeysOnePct}{34.5}  
\newcommand{\IntraBlockKeysTwoPct}{31.6}  
\newcommand{\IntraBlockKeysThreeToFivePct}{24.7}  
\newcommand{\IntraBlockKeysSixToTenPct}{6.3}  
\newcommand{\IntraBlockKeysGtTenPct}{2.9}  

\newcommand{\TotalContracts}{273,223}  
\newcommand{\ConcentrationTopOne}{16.5}  
\newcommand{\ConcentrationTopTwo}{31.4}  
\newcommand{\ConcentrationTopThree}{42.2}  
\newcommand{\ConcentrationTopTen}{63.6}  
\newcommand{\ConcentrationTopTwenty}{67.4}  
\newcommand{\ConcentrationTopTenThousand}{97.5}  

\newcommand{\GlobalTotalKeys}{35,035,731}  
\newcommand{\GlobalReuseFactor}{25.74}  
\newcommand{\GlobalKeysHundredPlus}{430,873}  
\newcommand{\GlobalKeysHundredPlusPct}{1.23}  
\newcommand{\GlobalAccessesHundredPlusPct}{84.0}  
\newcommand{\GlobalKeysOneToTwoPct}{64}  
\newcommand{\GlobalAccessesOneToTwoPct}{3.6}  

\newcommand{\LifespanKeysOneBlock}{25,382,837}  
\newcommand{\LifespanKeysTwoToFive}{7,497,511}  
\newcommand{\LifespanKeysSixToTwenty}{1,413,794}  
\newcommand{\LifespanKeysTwentyOneToFifty}{385,501}  
\newcommand{\LifespanKeysGtFifty}{356,088}  
\newcommand{\LifespanPctOneBlock}{72.4}  
\newcommand{\LifespanPctTwoToFive}{21.4}  
\newcommand{\LifespanPctSixToTwenty}{4.0}  
\newcommand{\LifespanPctTwentyOneToFifty}{1.1}  
\newcommand{\LifespanPctGtFifty}{1.0}  

\newcommand{\OverlapAvg}{12.4}  
\newcommand{\OverlapMedian}{11.7}  
\newcommand{\OverlapMax}{100.0}  

\newcommand{\UniqueKeysMedian}{1,806}  
\newcommand{\UniqueKeysPNinetyFive}{3,920}  
\newcommand{\UniqueKeysMax}{10,258}  

\newcommand{\TotalOpsMedian}{8,629}  
\newcommand{\TotalOpsPNinetyFive}{24,644}  
\newcommand{\TotalOpsMax}{109,627}  
\newcommand{\StorageOpsMedian}{6,587}  
\newcommand{\StorageOpsPNinetyFive}{19,058}  
\newcommand{\StorageOpsMax}{77,876}  
\newcommand{\PctBlocksLtTwoK}{58}  
\newcommand{\PctBlocksLtThreeK}{84}  
\newcommand{\PctBlocksGteFiveK}{1.2}  

\newcommand{\EvalBlockCount}{100,800}  
\newcommand{\EvalPrimaryExecTimeSec}{15,717}  
\newcommand{\EvalHintConstructTimeSec}{940}  
\newcommand{\EvalHintWriteTimeSec}{778}  
\newcommand{\EvalHintTotalTimeSec}{1,719}  
\newcommand{\EvalHintOverheadPct}{10.9}  
\newcommand{\EvalHintConstructOverheadPct}{6.0}  
\newcommand{\EvalHintWriteOverheadPct}{5.0}  
\newcommand{\EvalHintFractionMedianPct}{12.1}  
\newcommand{\EvalHintFractionPNinetyFivePct}{22.1}  
\newcommand{\EvalConstructMedianMs}{14.4}  
\newcommand{\EvalConstructPNinetyFiveMs}{33.8}  
\newcommand{\EvalWriteMedianMs}{12.5}  

\newcommand{\EvalConstructPerKeyUs}{8.1}  
\newcommand{\EvalSerializationMedianMs}{12.5}  
\newcommand{\EvalSerializationIQRLowMs}{11.9}  
\newcommand{\EvalSerializationIQRHighMs}{13.1}  
\newcommand{\EvalHintFracVarFromExecPct}{65}  
\newcommand{\EvalHintFracVarFromWritePct}{28}  
\newcommand{\EvalConstructKeysCorr}{0.79}  
\newcommand{\EvalWriteKeysCorr}{0.03}  
\newcommand{\EvalBinLtFiveHundredCount}{3,195}  
\newcommand{\EvalBinLtFiveHundredHintFracMedian}{24.8}  
\newcommand{\EvalBinLtFiveHundredExecMedianMs}{53}  
\newcommand{\EvalBinLtFiveHundredHintMedianMs}{13.0}  
\newcommand{\EvalBinFiveHundredToOneKCount}{4,787}  
\newcommand{\EvalBinFiveHundredToOneKHintFracMedian}{15.0}  
\newcommand{\EvalBinFiveHundredToOneKExecMedianMs}{123}  
\newcommand{\EvalBinFiveHundredToOneKHintMedianMs}{16.8}  
\newcommand{\EvalBinOneKToTwoKCount}{28,409}  
\newcommand{\EvalBinOneKToTwoKHintFracMedian}{12.7}  
\newcommand{\EvalBinOneKToTwoKExecMedianMs}{206}  
\newcommand{\EvalBinOneKToTwoKHintMedianMs}{24.8}  
\newcommand{\EvalBinTwoKToThreeKCount}{15,846}  
\newcommand{\EvalBinTwoKToThreeKHintFracMedian}{10.8}  
\newcommand{\EvalBinTwoKToThreeKExecMedianMs}{327}  
\newcommand{\EvalBinTwoKToThreeKHintMedianMs}{34.7}  
\newcommand{\EvalBinThreeKToFiveKCount}{6,753}  
\newcommand{\EvalBinThreeKToFiveKHintFracMedian}{9.8}  
\newcommand{\EvalBinThreeKToFiveKExecMedianMs}{440}  
\newcommand{\EvalBinThreeKToFiveKHintMedianMs}{42.5}  
\newcommand{\EvalBinGtFiveKCount}{194}  
\newcommand{\EvalBinGtFiveKHintFracMedian}{9.7}  
\newcommand{\EvalBinGtFiveKExecMedianMs}{536}  
\newcommand{\EvalBinGtFiveKHintMedianMs}{52.0}  

\newcommand{\EvalBaselineWallTimeSec}{23,250}  
\newcommand{\EvalPrimaryWallTimeSec}{29,889}  
\newcommand{\EvalPrimaryWallOverheadPct}{28.6}  

\newcommand{\EvalHintCompressedMedianKB}{46.5}  
\newcommand{\EvalHintCompressedMeanKB}{51.5}  
\newcommand{\EvalHintCompressedPNinetyFiveKB}{104.2}  
\newcommand{\EvalHintCompressedMaxKB}{239.0}  
\newcommand{\EvalHintRawMedianKB}{102.2}  
\newcommand{\EvalHintCompressionRatio}{2.17}  

\newcommand{\EvalBaselineTotalTimeSec}{22,604}  
\newcommand{\BaselineMedianMs}{203.0}  
\newcommand{\EvalSpeedupSequentialMedian}{24.9}  
\newcommand{\EvalSpeedupSequentialMean}{28.9}  
\newcommand{\EvalSpeedupSequentialPNinety}{45.0}  
\newcommand{\EvalSpeedupSequentialPNinetyNine}{103.0}  
\newcommand{\EvalSpeedupSequentialMax}{895.9}  
\newcommand{\EvalSlowerSequentialCount}{3,047}  
\newcommand{\EvalSlowerSequentialPct}{3.02}  
\newcommand{\EvalWallTimeSequentialSec}{4,366}  
\newcommand{\EvalWallSpeedupSequential}{5.2}  
\newcommand{\EvalSpeedupParallelSixteenMedian}{24.4}  
\newcommand{\EvalSpeedupParallelSixteenPNinety}{44.2}  
\newcommand{\EvalSpeedupParallelSixteenPNinetyNine}{100.3}  
\newcommand{\EvalSpeedupParallelSixteenMax}{750.7}  
\newcommand{\EvalSlowerParallelSixteenCount}{252}  
\newcommand{\EvalSlowerParallelSixteenPct}{0.25}  
\newcommand{\EvalWallTimeParallelSixteenSec}{956}  
\newcommand{\EvalWallSpeedupParallelSixteen}{23.6}  
\newcommand{\EvalSpeedupParallelSixtyFourMedian}{24.8}  
\newcommand{\EvalSpeedupParallelSixtyFourPNinety}{44.8}  
\newcommand{\EvalSpeedupParallelSixtyFourPNinetyNine}{102.8}  
\newcommand{\EvalSpeedupParallelSixtyFourMax}{856.6}  
\newcommand{\EvalSlowerParallelSixtyFourCount}{91}  
\newcommand{\EvalSlowerParallelSixtyFourPct}{0.09}  
\newcommand{\EvalWallTimeParallelSixtyFourSec}{871}  
\newcommand{\EvalWallSpeedupParallelSixtyFour}{25.9}  

\newcommand{\EvalScalingOneToSixteen}{4.6}  
\newcommand{\EvalScalingSixteenToSixtyFour}{1.1}  

\newcommand{\EvalBaselineRssGB}{26.3}  
\newcommand{\EvalPrimaryRssGB}{26.1}  
\newcommand{\EvalBackupRssGB}{26.0}  

\newcommand{\SourceTotalKeys}{200,828,943}  
\newcommand{\SourcePlainStateCount}{93,546,384}  
\newcommand{\SourceZeroCount}{21,016,404}  
\newcommand{\SourceChangesetCount}{86,266,155}  
\newcommand{\SourcePlainStatePct}{46.6}  
\newcommand{\SourceZeroPct}{10.5}  
\newcommand{\SourceChangesetPct}{43.0}  
\newcommand{\SourceNoHistoryPct}{57.0}  
\newcommand{\SourcePlainStateMedianPct}{47.5}  
\newcommand{\SourceZeroMedianPct}{5.6}  
\newcommand{\SourceChangesetMedianPct}{44.7}  
\newcommand{\SourceNoHistoryMedianPct}{55.3}  
\newcommand{\SourcePlainStatePTwentyFivePct}{43.5}  
\newcommand{\SourcePlainStatePSeventyFivePct}{50.9}  
\newcommand{\SourceZeroPTwentyFivePct}{4.0}  
\newcommand{\SourceZeroPSeventyFivePct}{9.2}  
\newcommand{\SourceChangesetPTwentyFivePct}{39.7}  
\newcommand{\SourceChangesetPSeventyFivePct}{48.6}  

\newcommand{\SecondWindowStorageOpsMedian}{3,894}  
\newcommand{\SecondWindowUniqueKeysMedian}{1,211}  
\newcommand{\SecondWindowBlockCount}{100,001}  
\newcommand{\SecondWindowHintOverheadPct}{7.3}  
\newcommand{\SecondWindowBaselineMedianMs}{45.3}  
\newcommand{\SecondWindowSpeedupParallelSixteenMedian}{9.0}  
\newcommand{\SecondWindowWallSpeedupParallelSixteen}{10.1}  
\newcommand{\SecondWindowSlowerParallelSixteenPct}{0.07}  
\newcommand{\SecondWindowStorageOpsRatio}{1.69}  
\newcommand{\SecondWindowUniqueKeysRatio}{1.49}  
\newcommand{\SecondWindowBaselineMsRatio}{4.5}  

\newcommand{\AdvHintBlockCount}{1,000}  
\newcommand{\AdvHintCapturedBlocks}{100,800}  
\newcommand{\AdvHintMedianKeys}{2,696}  
\newcommand{\AdvHintPrefetchThreads}{16}  
\newcommand{\AdvHintBaselineMedianMs}{41.1}  
\newcommand{\AdvHintMaliciousMedianMs}{63.5}  
\newcommand{\AdvHintSlowdownMedian}{1.58}  
\newcommand{\AdvHintSlowdownMean}{2.12}  
\newcommand{\AdvHintSlowdownPNinetyFive}{5.5}  
\newcommand{\AdvHintSlowdownPNinetyNine}{8.3}  
\newcommand{\AdvHintSlowerPct}{81.5}  

\begin{document}

\date{}

\title{\Large \bf Ira: Efficient Transaction Replay for Distributed Systems}

\author{
{\rm Adithya Bhat}\\
Visa Research
\and
{\rm Harshal Bhadreshkumar Shah}\\
Visa Research
\and
{\rm Mohsen Minaei}\\
Visa Research
} 

\maketitle

\begin{abstract}
  In primary-backup replication, the consensus latency is bounded by the time for backup nodes to replay (re-execute) transactions proposed by the primary.
Our key insight is that the primary, having already executed transactions, possesses knowledge of future access patterns which is the information needed for optimal replay by the backups.
In this work, we present Ira, a framework to accelerate backup replay by transmitting compact \emph{hints} alongside transaction batches.

We use Ethereum for our case study and present a concrete protocol, Ira-L, within our framework to improve cache management of Ethereum block execution.
The primaries implementing Ira-L provide hints that consist of the working set of keys used in an Ethereum block and one byte of metadata per key indicating the table to read from, and backups use these hints for efficient block replay.  

We evaluated Ira-L against the state-of-the-art Ethereum client reth over two weeks of Ethereum blocks ($100,800$ blocks, $24$ million transactions).
Our hint generation adds $10.9\%$ overhead to primary execution time.
On the backup, our hint-driven prefetching speeds up aggregate replay by $5.2\times$ with a single prefetch thread, and by $23.6\times$ with $16$ threads.
Our hints add a median of $47$ KB compressed metadata per block (${\sim}5\%$ of block payload).

\end{abstract}

\section{Introduction}
The primary-backup paradigm is the cornerstone of fault-tolerant distributed systems~\cite{grayDangersReplicationSolution1996}, from traditional databases~\cite{huntZooKeeperWaitfreeCoordination2010,junqueiraZabHighperformanceBroadcast2011} to modern blockchain networks~\cite{woodEthereumSecureDecentralised2014,yinHotStuffBFTConsensus2019}. In the standard State Machine Replication (SMR) model~\cite{schneiderImplementingFaulttolerantServices1990}, a designated \emph{primary} executes batches of transactions and propagates them to \emph{backup} replicas, who must replay these transactions to maintain consistent state.

We model each transaction as a sequence of read and write operations against a key-value store. The primary executes these operations, leveraging caches and indexes to optimize performance. When propagating the transaction batch to backups, the current systems transmit only the transaction data itself, forcing each backup to independently reconstruct the execution path. Transmitting all read/write values would incur prohibitive bandwidth, whereas replaying compact transaction logs allows backups to independently reconstruct the state.

SMR protocols~\cite{lamportPaxosMadeSimple2001,ongaroSearchUnderstandableConsensus2014} run a consensus protocol to ensure consistency of the state machine, but the latency of consensus depends on the response from a quorum of replicas (typically a majority or a super-majority). This makes replay critical to consensus performance.

This replay is inherently inefficient as the backups must make decisions (e.g., cache management) without knowledge of future access patterns, leading to sub-optimal eviction choices such as unnecessary waiting for I/O, redundant disk I/O, etc.

We illustrate this in \cref{fig:motivating}.
Consider a cache of size $2$ containing keys $\{A, B\}$, with upcoming accesses $C$, $A$, $C$. When $C$ arrives, we must evict one key. LRU evicts the least recently used key ($A$), causing a miss when we access $A$ next. An optimal strategy would be to evict $B$ avoiding the subsequent miss on $A$.
LRU incurs $2$ misses while optimal incurs only $1$.
Generalizing this, without knowledge of future accesses, backups using LRU or other heuristic policies cannot make these optimal decisions.

\begin{figure}[!htbp]
  \centering

\begin{tikzpicture}[
    scale=0.85,
    header/.style={font=\small\bfseries, minimum height=0.6cm},
    cell/.style={rectangle, draw, minimum width=2.0cm, minimum height=0.5cm,
                 font=\scriptsize, align=center},
    op/.style={cell, fill=gray!10},
    hit/.style={cell, fill=green!20},
    miss/.style={cell, fill=red!20},
    empty/.style={cell, fill=white},
    note/.style={font=\scriptsize\itshape, text=gray},
]

\def\colOp{0}
\def\colLRU{2.5}
\def\colOpt{5.0}

\def\rowHeader{0}
\def\rowInit{-0.8}
\def\rowOne{-1.6}
\def\rowTwo{-2.4}
\def\rowThree{-3.2}
\def\rowTotal{-4.2}

\node[header] at (\colOp, \rowHeader) {Operation};
\node[header] at (\colLRU, \rowHeader) {LRU Cache};
\node[header] at (\colOpt, \rowHeader) {Optimal Cache};

\draw[gray, dashed] (-1.1, \rowHeader-0.35) -- (6.1, \rowHeader-0.35);

\node[op] at (\colOp, \rowInit) {Initial};
\node[empty] at (\colLRU, \rowInit) {$\{A, B\}$};
\node[empty] at (\colOpt, \rowInit) {$\{A, B\}$};

\node[op] at (\colOp, \rowOne) {access $C$};
\node[miss] at (\colLRU, \rowOne) {$\{B, C\}$};
\node[miss] at (\colOpt, \rowOne) {$\{A, C\}$};
\node[note, right] at (\colOpt+1.1, \rowOne) {\textrm{evict} $B$};

\node[op] at (\colOp, \rowTwo) {access $A$};
\node[miss] at (\colLRU, \rowTwo) {$\{C, A\}$};
\node[hit] at (\colOpt, \rowTwo) {$\{A, C\}$};

\node[op] at (\colOp, \rowThree) {access $C$};
\node[hit] at (\colLRU, \rowThree) {$\{C, A\}$};
\node[hit] at (\colOpt, \rowThree) {$\{A, C\}$};

\draw[gray, dashed] (-1.1, \rowTotal+0.35) -- (6.1, \rowTotal+0.35);

\node[font=\scriptsize\bfseries] at (\colOp, \rowTotal) {Cache Misses};
\node[font=\scriptsize, text=red!70!black] at (\colLRU, \rowTotal) {2};
\node[font=\scriptsize, text=green!50!black] at (\colOpt, \rowTotal) {1};

\node[hit, minimum width=0.6cm] at (1.0, \rowTotal-0.8) {};
\node[font=\scriptsize, right] at (1.4, \rowTotal-0.8) {hit};
\node[miss, minimum width=0.6cm] at (3.2, \rowTotal-0.8) {};
\node[font=\scriptsize, right] at (3.6, \rowTotal-0.8) {miss};

\end{tikzpicture}
  \caption{LRU vs. optimal caching (cache size is $2$). For access sequence $C$, $A$, $C$ starting with cache $\{A, B\}$: LRU evicts $A$ (least recently used) when $C$ arrives, then misses when it accesses $A$ next. Optimal evicts $B$ (never accessed again), achieving one fewer miss. With hints revealing the access set, backups can prefetch all required keys before execution, eliminating misses entirely.}
  \label{fig:motivating}
\end{figure}

\subsection{Our Approach: Hint-Based Replay}

\begin{figure*}[!htbp]
  \centering
  \includegraphics[width=0.7\textwidth]{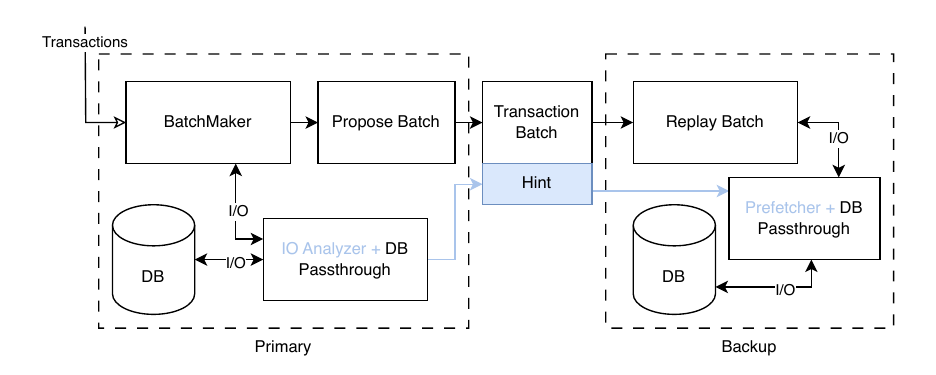}
  \caption{\sysname{} architecture. The primary executes transactions while building a \emph{hint} that records the access set. Backups use these hints to prefetch all required state before replay, achieving zero cache misses.}
  \label{fig:architecture}
\end{figure*}

When the cache is smaller than the working set, Belady's MIN algorithm~\cite{beladyStudyReplacementAlgorithms1966,mattsonEvaluationTechniquesStorage1970} provides the optimal eviction policy: evict the key whose next access is furthest in the future. However, when the cache can hold the entire working set, a simpler and more effective strategy is \emph{prefetching}: load all required keys before execution begins, eliminating misses entirely. In primary-backup replication, the primary already knows the complete access set---this knowledge enables prefetching on backups.

Our key insight is that the inefficiency from \cref{fig:motivating} stems from \emph{information asymmetry}: the primary possesses complete knowledge of the access pattern but discards this information when communicating with backups. The primary can encode this knowledge compactly and transmit it as \emph{hints} that enable backups to implement near-optimal replay. We propose \sysname{}, a framework to accelerate replay through hint-based replay. \Cref{fig:architecture} illustrates the architecture. \sysname{} operates two components:

\parhead{Primary} This component executes transactions while recording keys and their accesses, producing a hint. The concrete details of the hints depend on the execution and storage details of the application. This component is also responsible for delivering the hint to the backups.

\parhead{Backups} This component replays transactions using the hint to improve the storage access. This component is also responsible for receiving the hints from the primary.

\subsection{Contributions}

In summary, we make the following contributions:

\begin{itemize}[nosep]
  \item \textbf{Ethereum Case Study:} We systematically perform measurement studies on execution in Ethereum. We identify that I/O is a bottleneck, and caching plays a major role in improving the replay performance. 
  \item \textbf{Hint-Based Replay Architecture:} We design a system where the primary generates compact hints during execution that enable backups to implement optimal caching. Hints are advisory, i.e., incorrect hints affect performance but do not affect correctness, thus enabling incremental deployment.
  \item \parhead{\sysname{} for Ethereum} We present \sysname{}-L, which achieves zero-miss cache performance through per-block hints that enable complete working-set prefetching.
  \item \textbf{Performance evaluation:} We implement \sysname{} by modifying the state-of-the-art Rust Ethereum implementation (reth) and demonstrate $\EvalSpeedupSequentialMedian{}\times$ faster replay with only $\EvalHintOverheadPct{}\%$ hint generation overhead. With even a single backup node, \sysname{} achieves net speedup; benefits grow with additional replicas.
\end{itemize}

\section{Model and Preliminaries}
\label{sec:prelim}
\parhead{Cache Optimization with Future Knowledge}
Belady's MIN algorithm~\cite{beladyStudyReplacementAlgorithms1966}, proved optimal by Mattson et al.~\cite{mattsonEvaluationTechniquesStorage1970}, provides the theoretically optimal cache replacement policy: when eviction is necessary, evict the item whose next access is \emph{furthest in the future}. However, when the cache can hold the entire working set, a simpler strategy---\emph{prefetching}---is more effective: load all required keys before execution begins, eliminating misses entirely. Both strategies require knowledge of future accesses, information that is typically unavailable but is available to the primary in primary-backup replication~(\Cref{subsec:ira-architecture}).

\subsection{Primary-backup system model}

We assume the standard primary-backup replication model.

\parhead{Primary Node}
The primary executes transactions against a key-value store (e.g., MDBX) and proposes a batch of transactions for replication. 

\parhead{Backup Nodes}
Backups receive transaction batches and associated metadata from the primary. The backups replay the transactions with standard configured cache policies (e.g., LRU). Backups independently verify execution results, ensuring correctness of primary's proposal.

\parhead{Execution Flow}
A transaction batch proceeds through the system as follows:
\begin{enumerate}[nosep]
  \item The primary executes the batch, generating state updates.
  \item The primary commits the batch to durable storage and transmits the log to backups.
  \item Each backup replays the batch.
  \item Backups acknowledge replay completion to the primary.
\end{enumerate}

\subsection{Ethereum: A Primer}%
\label{subsec:eth primer}

Ethereum~\cite{woodEthereumSecureDecentralised2014} is a decentralized blockchain platform that executes arbitrary programs called \emph{smart contracts}. The execution client runs the Ethereum Virtual Machine (EVM), a stack-based virtual machine that interprets contract bytecode. We provide background on the aspects relevant to understanding \sysname{}-L\iffullversion; \Cref{sec:reth arch} describes the execution client architecture in detail\fi.

\parhead{Client Architecture}
Ethereum nodes run two cooperating clients: a \emph{consensus client} that participates in the proof-of-stake protocol to determine the canonical chain, and an \emph{execution client} that executes transactions and maintains state. The consensus client receives blocks from the network and instructs the execution client to validate them by re-executing all transactions and verifying the resulting \emph{state root}, the cryptographic commitment to all account state.

\parhead{Blocks and Transactions}
A block contains an ordered sequence of transactions. Each transaction specifies a sender, recipient, value transfer, and optional calldata for smart contract invocation. The execution client processes transactions sequentially: each transaction observes the state modifications of all preceding transactions in the same block. This deterministic execution model ensures that all nodes converge to identical state.

\parhead{State Model}
Ethereum state is a mapping from $20$-byte \emph{addresses} to \emph{accounts}. Each account contains:

\italhead{Balance} The account's Ether holdings (256-bit integer)

\italhead{Nonce} A counter incremented with each outgoing transaction

\italhead{Code} For contract accounts, the immutable bytecode (up to $24$ KB)

\italhead{Storage} For contract accounts, a mapping from $32$-byte slots to $32$-byte values

Contract storage is the dominant source of state I/O: a single contract may have millions of storage slots, and complex DeFi protocols access hundreds of slots per transaction.

The EVM is a stack-based virtual machine that executes contract bytecode. State-accessing opcodes include:

\texttt{SLOAD}/\texttt{SSTORE}: Read/write contract storage slots

\texttt{BALANCE}/\texttt{SELFBALANCE}: Query account balances

\texttt{CALL}/\texttt{STATICCALL}/\texttt{DELEGATECALL}: Invoke other contracts (loads bytecode)

\texttt{EXTCODESIZE}/\texttt{EXTCODECOPY}: Query contract bytecode

Each state-accessing opcode translates to one or more database lookups in the execution client. \Cref{sec:ethereum case study} characterizes the distribution and locality of these operations.

\parhead{Block Validation as Primary-Backup Replay}
In Ethereum's proof-of-stake consensus, a designated \emph{proposer} (analogous to the primary) constructs and executes a new block, then broadcasts it to \emph{validators} (analogous to backups). Validators must re-execute the block to verify correctness before attesting to it. This replay is the bottleneck that \sysname{}-L addresses.

\subsection{\sysname{}: Hint-Based Replay Architecture}%
\label{subsec:ira-architecture}

\sysname{} addresses the information asymmetry between primary and backups by having the primary encode its access pattern as a compact \emph{hint} and transmit it alongside the transaction batch. The hint contains sufficient information for backups to prefetch required state before replay, eliminating cache misses. Critically, hints are \emph{advisory}: they affect only performance, not correctness. If a hint is missing, corrupted, or incomplete, the backup falls back to standard replay without prefetching.

\parhead{Protocol Overview}
The \sysname{} protocol operates in five phases:
\begin{enumerate}[nosep]
  \item \textbf{Primary Execution:} The primary executes transactions while instrumenting key accesses, recording each key read or written into an \emph{access set} $A$.
  \item \textbf{Hint Construction:} After execution, the primary constructs a hint $H$ containing the access set $A$, optionally annotated with metadata (e.g., where to read each key, or access ordering if cache capacity is limited).
  \item \textbf{Hint Transmission:} The primary transmits $H$ alongside (or ahead of) the transaction batch to backups.
  \item \textbf{Backup Prefetch:} Backups use $H$ to prefetch all keys in $A$ into their cache before execution begins.
  \item \textbf{Backup Replay:} Backups execute transactions with all required state pre-cached, achieving near-zero I/O latency during execution.
\end{enumerate}

\parhead{Hint Contents}
At minimum, a hint contains the \emph{access set} $A$: the set of unique keys accessed during batch execution. Richer hints may include additional metadata:
\begin{itemize}[nosep]
  \item \textbf{Source annotations:} Where to read each key (e.g., the \emph{main state table} that holds each key's current value vs.\ \emph{historical change sets} that record prior values), enabling the backup to route reads to the fastest storage path.
  \item \textbf{Access ordering:} The sequence in which keys are accessed, enabling optimal cache eviction when the access set exceeds cache capacity.
  \item \textbf{Read values:} The actual values read by the primary, trading bandwidth for backup computation when network capacity exceeds storage throughput.
\end{itemize}

The choice of hint granularity depends on the system's I/O characteristics, network bandwidth, and memory constraints. \Cref{sec:ira-l} presents \sysname{}-L, a concrete instantiation that uses per-block access sets with source annotations for Ethereum block replay.
\iffullversion
\Cref{sec:appendix:general-ira} discusses how to adapt \sysname{} to other primary-backup systems including Raft-based databases, streaming replication, and distributed key-value stores.
\fi

\parhead{Threat Model and Security Goals}
\label{subsec:threat-model}
\sysname{} is a performance optimization assuming that the underlying replication protocol provides consensus ensuring safety and liveness.
Our system \sysname{} inherits these guarantees.
Our hints are \emph{advisory} and affect only replay latency.

\italhead{Trust} We assume (i)~the replication protocol is secure, i.e., ensures safety and liveness, and (ii)~backups verify execution outputs against the primary's commitment, e.g., the Ethereum state root~(\Cref{sec:correctness}). The primary may be Byzantine.

\italhead{Adversary} The adversary controls the primary and any nodes the underlying protocol tolerates such as the clients, and schedules hint and block delivery within the protocol's network assumptions.
Against \sysname{} specifically, it may (a)~omit hints, (b)~send a hint that fails an integrity check, (c)~send a valid hint that omits required keys, or (d)~send a hint with spurious keys. Our security goal is that none of these cause a backup to accept an incorrect result or to diverge from the state it would compute without hints; a malicious hint may affect only replay latency. \Cref{sec:correctness} and~\Cref{sec:discussion} establish how \sysname{}-L detects and absorbs each case.

\italhead{Security implications} Faster backup replay carries positive security implications, e.g., a weakened verifier's dilemma and improved denial-of-service resilience, which we discuss in~\Cref{sec:discussion}.

\italhead{Out of scope} Hints reveal storage-access patterns, public on Ethereum but unsuited to permissioned settings with confidential transactions. Identifying and penalizing peers that repeatedly deliver invalid hints, incentivizing correct hints, and compute-bound replay are deployment concerns we treat in~\Cref{sec:discussion}.

\section{Ethereum: A case study}
\label{sec:ethereum case study}
Before designing \sysname{}-L, we conduct a systematic characterization of Ethereum state access patterns to understand the structure and locality of block execution workloads.

\subsection{Methodology}

To characterize Ethereum state access patterns, we re-executed two weeks of historical mainnet blocks and recorded every state operation performed by the EVM. Re-execution against archival state ensures deterministic, reproducible traces that capture the exact sequence of operations from original block validation.

\parhead{Instrumentation}
We instrumented \emph{reth}~\cite{ParadigmxyzReth2026}, a production Rust Ethereum client, using its EVM Inspector interface. The Inspector intercepts every EVM instruction during execution, allowing us to capture state operations at the exact moment they occur. For each state-accessing opcode, we perform \emph{stack introspection} to extract operand values (target addresses, storage slots) directly from the EVM stack before the operation executes.

We captured $14$ distinct operation types: storage operations (\texttt{SLOAD}, \texttt{SSTORE}), account metadata queries (\texttt{BALANCE}, \texttt{SELFBALANCE}, \texttt{EXTCODESIZE}, \texttt{EXTCODEHASH}), bytecode access (\texttt{CALL}, \texttt{CALLCODE}, \texttt{STATICCALL}, \texttt{DELEGATECALL}, \texttt{EXTCODECOPY}), and state modifications (\texttt{CREATE}, \texttt{CREATE2}, \texttt{SELFDESTRUCT}). \Cref{tab:trace-schema} shows the trace database schema.

\begin{table}[!ht]
\centering
\caption{Trace database schema for EVM state operations.}
\label{tab:trace-schema}
\small
\begin{tabular}{lll}
\toprule
\textbf{Field} & \textbf{Type} & \textbf{Description} \\
\midrule
\texttt{block\_number} & uint64 & Block containing operation \\
\texttt{tx\_index} & uint16 & Transaction position in block \\
\texttt{op\_index} & uint32 & Operation position in transaction \\
\texttt{op\_type} & uint8 & Operation type (14 types) \\
\texttt{target\_address} & bytes20 & Contract or account accessed \\
\texttt{storage\_slot} & bytes32 & Storage slot (nullable) \\
\texttt{value\_size} & uint32 & Bytecode size (nullable) \\
\bottomrule
\end{tabular}
\end{table}

\parhead{Block Processing}
For each block, we retrieve the parent block's state from reth's archival database and execute all transactions sequentially. To ensure correct intra-block state visibility, we maintain a \emph{write-through cache} that accumulates state modifications across transactions within a block: each transaction observes the effects of all preceding transactions, matching original execution semantics. Blocks are processed in parallel across CPU cores, with results merged and sorted by $(block, tx, op)$ order.

\parhead{Dataset}
We collected traces from blocks $24,019,447$ through $24,120,246$ ($100,800$ consecutive blocks spanning December 15--29, 2025 on Ethereum mainnet). The trace contains $\OpsGrandTotal{}$ state operations across approximately $15$ million transactions. We store traces in Apache Parquet~\cite{apachesoftwarefoundationApacheParquet} columnar format with Zstandard~\cite{facebookZstandardRealtimeData} compression, yielding a ${\sim}12$ GB dataset.

\parhead{Completeness}
Our instrumentation captures $100\%$ of state operations executed during block validation, with no sampling or filtering. Transactions that fail during re-execution (e.g., due to state inconsistencies) are excluded; these represent less than $0.01\%$ of transactions and do not affect our findings.

\italhead{Implementation Note}
Reth uses flat storage (the \texttt{PlainStorageState} table in MDBX, a high-performance key-value store) for execution-time state access, rather than direct Merkle Patricia Trie traversal as specified by the Ethereum Yellow Paper~\cite{woodEthereumSecureDecentralised2014}. Each \texttt{SLOAD} or \texttt{SSTORE} corresponds to a single key-value lookup. Our traces capture these logical operations, which directly correspond to database I/O during block execution.

\subsection{Findings}

\begin{figure}[!ht]
  \centering
  \includegraphics[width=\columnwidth]{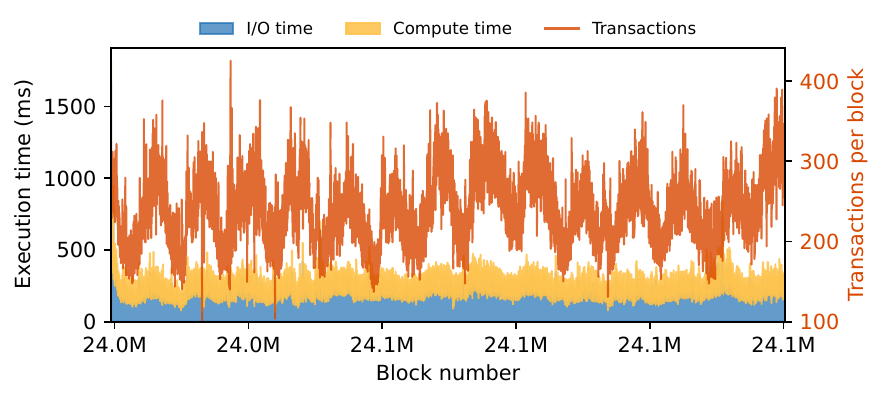}
  \caption{Execution time breakdown showing I/O dominance. The stacked area shows I/O time (blue) and compute time (yellow) per block; the line shows transaction count. I/O consistently dominates regardless of block size.}
  \label{fig:io-breakdown}
\end{figure}

\parhead{\#1. I/O dominates execution}
State I/O overwhelmingly dominates block execution time rather than EVM computation. \Cref{fig:io-breakdown} shows the breakdown of execution time across $\IoTraceBlockCount{}$ consecutive blocks: I/O accounts for $\IoTimePct{}\%$ of total execution time, while pure EVM computation consumes only $\ComputeTimePct{}\%$. The median per-block I/O share is $\MedianIoTimePct{}\%$, with the $5$-$95$th percentile range spanning $\IoTimePctPFive{}\%$--$\IoTimePctPNinetyFive{}\%$, indicating consistent I/O dominance across blocks regardless of transaction volume. We attribute to I/O the time spent inside backend state-access calls (account, storage, bytecode, and block-hash reads); all remaining time, including EVM opcode execution and accesses served from the in-memory cache, counts as compute.

This $\IoComputeRatio{}$:$1$ of I/O-to-compute ratio has a critical implication: optimizing state access patterns yields far greater performance gains than optimizing EVM execution. Reduction in I/O latency through caching, prefetching, or sequential access patterns can significantly accelerate block validation.

\parhead{\#2. Storage dominates I/O}
\Cref{tab:op-distribution} shows the distribution of state operations in our trace\iffullversion{} (\Cref{fig:op-distribution} in the appendix shows the per-block breakdown)\fi. Storage operations (\texttt{SLOAD} and \texttt{SSTORE}) account for $\StorageOpsPct{}\%$ of all state accesses. Call operations, while numerous, contribute minimal I/O overhead: contract bytecode is loaded once per contract and cached in memory, so repeated calls to the same contract incur no disk I/O. Account metadata queries comprise less than $1\%$ of operations.

\begin{table}[t]
\centering
\caption{State operation distribution across $100,800$ blocks.}
\label{tab:op-distribution}
\small
\begin{tabular}{lrr}
\toprule
\textbf{Category} & \textbf{Operations} & \textbf{Share} \\
\midrule
Storage (\texttt{SLOAD}/\texttt{SSTORE}) & $\StorageOpsTotal{}$ & $\StorageOpsPct{}\%$ \\
Calls (bytecode load) & $\CallsTotal{}$ & $\CallsPct{}\%$ \\
Account metadata & $\AccountMetadataTotal{}$ & $\AccountMetadataPct{}\%$ \\
Creation/destruction & $\CreationDestructionTotal{}$ & $\CreationDestructionPct{}\%$ \\
\midrule
\textbf{Total} & $\OpsGrandTotal{}$ & $100\%$ \\
\bottomrule
\end{tabular}
\end{table}

The read-write ratio for storage operations is $\StorageRwRatio{}:1$ ($\SloadCount{}$ \texttt{SLOAD}s vs.\ $\SstoreCount{}$ \texttt{SSTORE}s), indicating that reads dominate the workload. Since reads benefit directly from cache hits while writes must persist regardless, this ratio suggests significant caching potential.

\parhead{\#3. High intra-block locality}
We define the \emph{reuse factor} as the ratio of total storage operations to unique storage keys accessed. Globally, the reuse factor is $\GlobalReuseFactor{}\times$ ($\StorageOpsTotal{}$ operations across $\GlobalTotalKeys{}$ unique keys). However, this conflates intra-block and cross-block reuse.

\begin{figure}[t]
  \centering
  \includegraphics[width=\columnwidth]{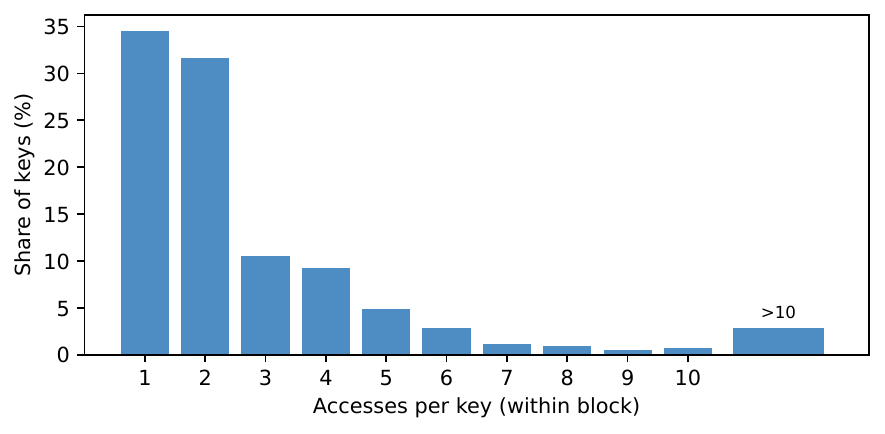}
  \caption{Intra-block access frequency distribution. Most keys are accessed 1--2 times per block, but a long tail of frequently-accessed keys drives high locality.}
  \label{fig:intra-block-frequency}
\end{figure}

To isolate intra-block locality, we examine the distribution of per-key access counts within blocks. \Cref{fig:intra-block-frequency} shows this distribution: $\IntraBlockSingleAccessPct{}\%$ of keys are accessed only once within their block, while $\IntraBlockMultiAccessPct{}\%$ are accessed multiple times. The distribution exhibits a long tail, with some keys accessed up to $\IntraBlockMaxAccessCount{}$ times within a single block.

This finding has important implications: we observe that $\IntraBlockRepeatedAccessPct{}\%$ of key accesses within a block are to keys already accessed earlier in that block (i.e., of $\IntraBlockTotalAccesses{}$ storage accesses, only $\IntraBlockTotalKeys{}$ are first accesses). A cache warmed with the correct keys before execution begins can serve most accesses as hits, thus underscoring the importance of an efficient caching system to improve block execution.

\begin{figure}[!htb]
  \centering
  \includegraphics[width=\columnwidth]{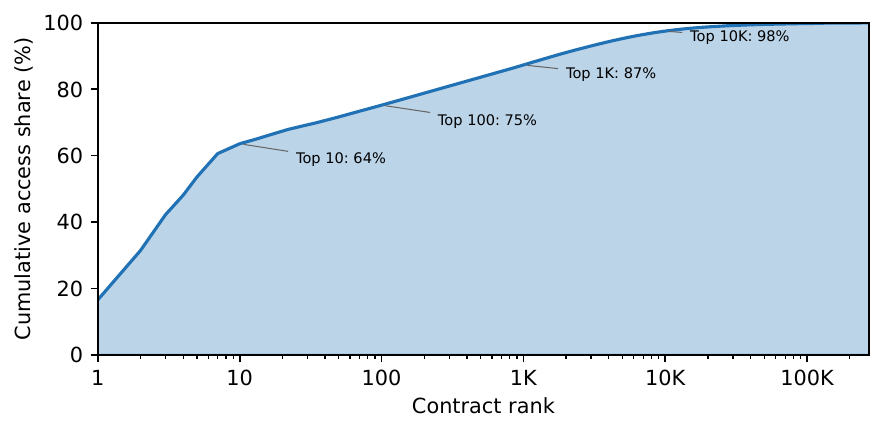}
  \caption{Cumulative storage access concentration by contract rank (out of $\TotalContracts{}$ total contracts). A small number of contracts dominate state access: the top $10$ contracts account for $\ConcentrationTopTen{}\%$ and the top $10{,}000$ account for $\ConcentrationTopTenThousand{}\%$ of all storage operations.}
  \label{fig:contract-concentration}
\end{figure}

\parhead{\#4. Extreme Access Concentration}
Access patterns exhibit extreme concentration at both the contract and key levels. \Cref{fig:contract-concentration} shows the cumulative storage access share as a function of contract rank: the top $3$ contracts (USDT, USDC, and Compound) account for $\ConcentrationTopThree{}\%$ of all storage operations, and the top $10$ contracts account for $\ConcentrationTopTen{}\%$.\iffullversion{} \Cref{tab:concentration} in the appendix provides the detailed breakdown.\fi

At the key level, concentration is even more pronounced. The top $\GlobalKeysHundredPlusPct{}\%$ of keys ($\GlobalKeysHundredPlus{}$ keys with $100+$ accesses) account for $\GlobalAccessesHundredPlusPct{}\%$ of all storage operations. Conversely, $\GlobalKeysOneToTwoPct{}\%$ of keys are accessed only once or twice, contributing just $\GlobalAccessesOneToTwoPct{}\%$ of total accesses.
This concentration might suggest that caching the most frequently accessed keys globally would suffice. However, as Finding~\#5 shows, $\LifespanPctOneBlock{}\%$ of keys are ephemeral (appearing in only one block), and consecutive-block overlap averages just $\OverlapAvg{}\%$. The ``hot'' keys change unpredictably from block to block, making static caching ineffective.

\begin{figure}[!ht]
  \centering
  \includegraphics[width=\columnwidth]{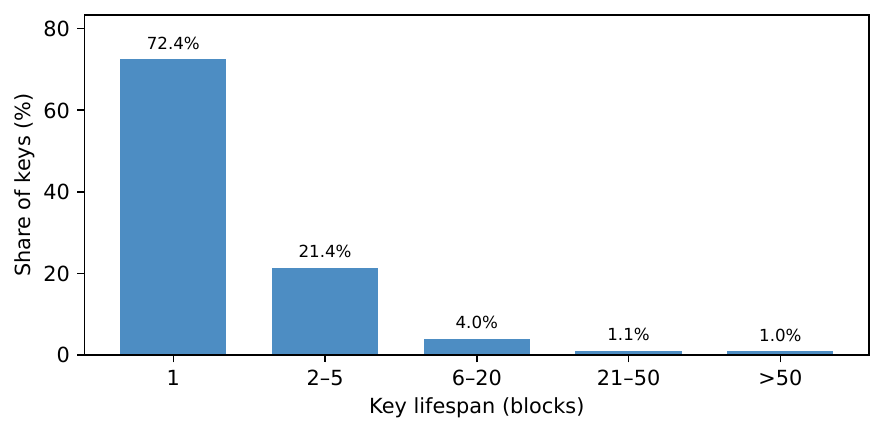}
  \caption{Key lifespan distribution: most keys are ephemeral and appear in only one block ($\LifespanPctOneBlock{}\%$), while less than $\LifespanPctGtFifty{}\%$ persist across more than $50$ blocks.}
  \label{fig:key-lifespan}
\end{figure}

\parhead{\#5. Ephemeral Key Dominance}
Despite high intra-block reuse, most keys are ephemeral and appear in only one block. \Cref{fig:key-lifespan} shows the distribution of key lifespans: $\LifespanPctOneBlock{}\%$ of keys appear in only a single block, and $\LifespanPctTwoToFive{}\%$ appear in $2$--$5$ blocks. Only $\LifespanPctGtFifty{}\%$ of keys persist across more than $50$ blocks.\iffullversion{} \Cref{tab:key-lifespan} in the appendix provides the detailed breakdown.\fi

Consecutive block overlap, i.e., the fraction of keys in block $n+1$ that also appeared in block $n$, averages only $\OverlapAvg{}\%$, has a median of $\OverlapMedian{}\%$, and a max of $\OverlapMax{}\%$. This low overlap implies that standard cache replacement policies like LRU cannot effectively predict which keys will be needed in upcoming blocks: LRU retains recently-used keys, but with only $\OverlapAvg{}\%$ overlap, most cached keys will not be accessed in the next block, while most keys needed will not be in cache.

\begin{figure}[!ht]
  \centering
  \includegraphics[width=\columnwidth]{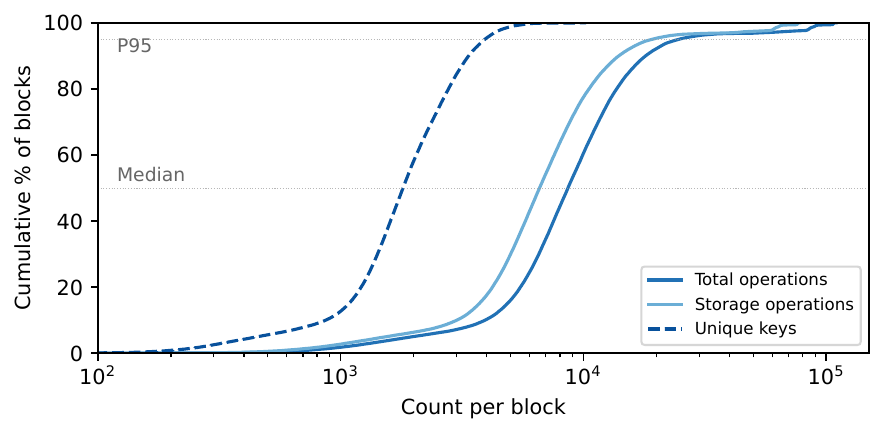}
  \caption{Cumulative distribution of per-block statistics. The working set remains bounded: $50\%$ of blocks access fewer than $\UniqueKeysMedian{}$ unique keys, and $95\%$ access fewer than $\UniqueKeysPNinetyFive{}$.}
  \label{fig:per-block-stats}
\end{figure}

\parhead{\#6. Bounded Working Set}
\Cref{fig:per-block-stats} summarizes per-block statistics. The median block performs $\TotalOpsMedian{}$ total operations, of which $\StorageOpsMedian{}$ are storage operations accessing $\UniqueKeysMedian{}$ unique keys. Even at the $95$th percentile, blocks access only $\UniqueKeysPNinetyFive{}$ unique keys.\iffullversion{} \Cref{tab:per-block} in the appendix provides the detailed breakdown.\fi

This bounded working set implies that per-block hints are tractable: hinting up to $2{,}000$ keys requires at most $101$ KB uncompressed ($52$ bytes per key: $20$-byte address + $32$-byte slot), which can be compressed significantly. The majority of blocks ($\PctBlocksLtTwoK{}\%$) access fewer than $2{,}000$ unique keys, and $\PctBlocksLtThreeK{}\%$ access fewer than $3{,}000$. Only $\PctBlocksGteFiveK{}\%$ of blocks exceed $5{,}000$ unique keys. This tight distribution suggests that a fixed-size hint buffer of $2{,}000$--$3{,}000$ keys would cover the vast majority of blocks, making per-block prefetching hints both practical and efficient.

\italhead{Takeaway}
Our analysis reveals a fundamental tension. On one hand, I/O dominates execution ($\IoTimePct{}\%$), storage dominates I/O ($\StorageOpsPct{}\%$), and per-block working sets are bounded (median $\UniqueKeysMedian{}$ keys), making caching critical and per-block hints tractable. On the other hand, $\LifespanPctOneBlock{}\%$ of keys are ephemeral and consecutive-block overlap averages just $\OverlapAvg{}\%$, so LRU and other heuristic policies cannot predict which keys the next block will need. \sysname{}-L~(\Cref{sec:ira-l}) closes this gap by transmitting the primary's access set to backups.

\section{\sysname{}-L: Latency-Optimal Algorithm}
\label{sec:ira-l}
\sysname{}-L instantiates the \sysname{} framework~(\Cref{subsec:ira-architecture}) for Ethereum block replay.
After executing block~$b$, the primary constructs a \emph{hint} $H_b$ containing three components: storage keys with source annotations, account addresses, and bytecode addresses (defined in \Cref{sec:hint-generation}).
The primary transmits $H_b$ alongside the block data.
The backup uses $H_b$ to prefetch all required state \emph{before} execution begins, eliminating cold misses entirely.

We define the cold-miss barrier formally.
Let $A_b$ denote the set of unique storage keys accessed during execution of block~$b$.
Without hints, the backup discovers each key in $A_b$ only at the moment it is first referenced, incurring I/O latency on every first access.
Because I/O dominates execution time ($\IoComputeRatio{}$:$1$ I/O-to-compute ratio), eliminating these misses through prefetching reduces backup execution time toward the pure-compute lower bound.
\Cref{sec:eval} confirms that backup execution closely approaches this bound, achieving a median $\EvalSpeedupSequentialMedian{}\times$ per-block speedup over the baseline.

\subsection{Hint Generation}
\label{sec:hint-generation}

\parhead{Key Collection}
During execution of block~$b$, an EVM inspector intercepts all state-accessing opcodes and records each accessed key into three sets:
\begin{itemize}[nosep]
  \item \emph{Storage keys}: $(\mathit{address}, \mathit{slot})$ pairs from \texttt{SLOAD} and \texttt{SSTORE}.
  \item \emph{Account addresses}: from \texttt{BALANCE}, \texttt{SELFBALANCE}, \texttt{EXTCODESIZE}, and \texttt{EXTCODEHASH}.
  \item \emph{Bytecode addresses}: from \texttt{CALL}, \texttt{STATICCALL}, \texttt{DELEGATECALL}, \texttt{EXTCODECOPY}, \texttt{CREATE}, and \texttt{CREATE2}.
\end{itemize}
Additionally, transaction senders, recipients, and the block beneficiary (coinbase) are added to the account set, since the EVM accesses these accounts implicitly during fee processing and reward distribution.
The inspector is lightweight: it reads stack values at each relevant opcode without modifying execution behavior.

\parhead{Source Annotation}
For each storage key $(a, k)$, the primary determines its \emph{source}, which is a single-byte annotation indicating where the backup should read the value:
\begin{itemize}[nosep]
  \item \textbf{PlainState} ($\sigma = 0$): The current value resides in the main state table. A single B-tree seek suffices.
  \item \textbf{Zero} ($\sigma = 1$): The slot was never written. The value is zero. No disk I/O is required.
  \item \textbf{Changeset} ($\sigma = 2$): The value must be read from historical change sets. The full two-step lookup is required.
\end{itemize}

Without source annotations, the backup must perform a two-step lookup for every key: first seeking the history index to locate the value, then reading it from the appropriate table.
The source annotation eliminates the first step for Plain State keys, reducing two seeks to one and eliminating all I/O for Zero keys.
Across our dataset, $\SourcePlainStatePct{}\%$ of storage keys resolve to PlainState, $\SourceZeroPct{}\%$ to Zero, and $\SourceChangesetPct{}\%$ to Changeset\iffullversion{} (\Cref{sec:appendix:source-dist})\fi.
In total, $\SourceNoHistoryPct{}\%$ of keys bypass the history index entirely, with a per-block median of $\SourceNoHistoryMedianPct{}\%$.
The source hint removes the confirmation step for PlainState keys and all I/O for Zero keys; only Changeset keys require the full two-step lookup.
\iffullversion\Cref{sec:reth arch} describes the full history-index lookup algorithm.\fi

\parhead{Serialization}
The hint is serialized in a compact binary format: each storage key requires $53$ bytes ($20$-byte address $+$ $32$-byte slot $+$ $1$-byte source annotation), and each address key requires $20$ bytes. Addresses and slots are fixed-width big-endian byte arrays, so the format is byte-order-explicit and portable across client, OS, and ISA.
The serialized hint is compressed with Zstandard~\cite{facebookZstandardRealtimeData} and stored in a key-value database indexed by block number.
\Cref{sec:eval} reports that the median compressed hint is $\EvalHintCompressedMedianKB{}$ KB, achieving a $\EvalHintCompressionRatio{}\times$ compression ratio which is negligible relative to the ${\sim}2$ MB block payload.


\begin{algorithm}[t]
\caption{\sysname{}-L Hint Generation (Primary)}
\label{alg:ira-primary}
\begin{algorithmic}[1]
\Require Block $B = \langle T_1, \ldots, T_n \rangle$, State database $\mathcal{D}$
\Ensure Hint set $H_B$ for block $B$
\State $S \gets \emptyset$; $A \gets \emptyset$; $C \gets \emptyset$ \Comment{Storage, accounts, code}
\LineComment{Phase 1: Execute and collect accessed keys}
\For{$i \gets 1$ \textbf{to} $n$}
    \State Execute $T_i$ with instrumented EVM
    \State $S \gets S \cup \{(a, k) : T_i \text{ accesses storage slot } k \text{ of account } a\}$
    \State $A \gets A \cup \{a : T_i \text{ accesses balance or nonce of } a\}$
    \State $C \gets C \cup \{a : T_i \text{ accesses bytecode of } a\}$
\EndFor
\LineComment{Phase 2: Annotate storage keys with source information}
\For{$(a, k) \in S$}
    \State $b^* \gets$ largest $b < B.\mathit{num}$ where $(a, k)$ was modified, or $\bot$
    \If{$b^* = \bot$}
        \State $\mathit{src} \gets \textsc{Zero}$ \Comment{Slot never written; value is zero}
    \ElsIf{$b^*$ is in pruned history}
        \State $\mathit{src} \gets \textsc{Plain}$ \Comment{Read from plain state table}
    \Else
        \State $\mathit{src} \gets \textsc{Hist}$ \Comment{Read from historical changeset}
    \EndIf
    \State $S[a, k] \gets (a, k, \mathit{src})$
\EndFor
\LineComment{Phase 3: Serialize and compress}
\State $H_B \gets \textsc{Compress}(S, A, C)$
\State \Return $H_B$
\end{algorithmic}
\end{algorithm}

\Cref{alg:ira-primary} summarizes the hint generation procedure.
The primary executes each transaction with the inspector attached (Phase~1), annotates every storage key with its optimal source (Phase~2), and serializes the result (Phase~3).

\subsection{Hint-Based Replay}
\label{sec:hint-replay}

\parhead{Sorted Batch Prefetching}
Our key algorithmic contribution on the backup side is \emph{sorted batch prefetching}.
Upon receiving hints for a batch of blocks, the backup proceeds as follows:
\begin{enumerate}[nosep]
  \item Collects all storage keys from the batch into a single sorted collection, ordered lexicographically by $(\mathit{address}, \mathit{slot})$.
  \item Routes keys by source annotation: PlainState keys are read via a single forward cursor walk through the B-tree, converting random I/O into a sequential scan; Zero keys are resolved immediately to zero with no I/O; Changeset keys are deferred to on-demand lookup during execution.
  \item Similarly loads account metadata and bytecodes via sorted cursor scans.
\end{enumerate}

Sorting is critical because the underlying database (MDBX, an LMDB-derived B-tree store) stores data in sorted key order.
Sorted access maximizes page-cache locality: the cursor advances monotonically through the B-tree, touching each page at most once.
Without sorting, each random seek may traverse $O(\log N)$ pages.
By batching keys from multiple blocks (default batch size: $32$ blocks), the amortized cost per key approaches a single sequential read.

\parhead{Pipelined Execution}
The backup operates in two phases:
\begin{itemize}[nosep]
  \item \emph{Phase~1 (warm-up):} Parallel prefetch of approximately $3{,}000$ blocks into an $8$ GB in-memory buffer, using a thread pool for initial catch-up.
  \item \emph{Phase~2 (steady-state):} A dedicated prefetcher thread reads hints and prefetches state in sorted batches, feeding populated caches to the executor through a bounded channel. The executor processes blocks sequentially, consuming prefetched state from the channel. The batch size is tuned so that the next batch completes before the executor finishes the current one, hiding prefetch latency behind execution.
\end{itemize}

This pipelining ensures that the backup's wall-clock time is bounded by $\max(\text{prefetch}, \text{execution})$ rather than their sum.
\iffullversion\Cref{sec:reth arch} provides implementation details of the threading model and channel sizing.\fi

\parhead{Memory Discipline}
Each block's cache is discarded after execution and there is no cross-block cache.
Since hints provide complete coverage for each block, retaining state across blocks is unnecessary.
This bounds memory usage to a single block's working set and avoids the complexity of LRU eviction.


\begin{algorithm}[t]
\caption{\sysname{}-L Replay (Backup)}
\label{alg:ira-backup}
\begin{algorithmic}[1]
\Require Block $B = \langle T_1, \ldots, T_n \rangle$, Hints $H_B$, State database $\mathcal{D}$
\Ensure Execution matches primary
\State $(S, A, C) \gets \textsc{Decompress}(H_B)$
\State $\mathit{cache} \gets \emptyset$
\LineComment{Phase 1: Sort keys for sequential database access}
\State $S' \gets \textsc{SortLexicographic}(S)$ \Comment{Sort by $(a, k)$}
\State $A' \gets \textsc{Sort}(A)$; \quad $C' \gets \textsc{Sort}(C)$
\LineComment{Phase 2: Prefetch state via sequential cursor traversal}
\For{$a \in A'$} \Comment{Sequential scan}
    \State $\mathit{cache}[a] \gets \mathcal{D}.\textsc{ReadAccount}(a)$
\EndFor
\For{$(a, k, \mathit{src}) \in S'$} \Comment{Sequential scan}
    \If{$\mathit{src} = \textsc{Zero}$}
        \State $\mathit{cache}[a, k] \gets 0$ \Comment{No I/O required}
    \ElsIf{$\mathit{src} = \textsc{Plain}$}
        \State $\mathit{cache}[a, k] \gets \mathcal{D}.\textsc{ReadPlain}(a, k)$
    \Else
        \State $\mathit{cache}[a, k] \gets \mathcal{D}.\textsc{ReadHist}(a, k, B.\mathit{num})$
    \EndIf
\EndFor
\For{$a \in C'$}
    \State $\mathit{cache}[a.\mathit{code}] \gets \mathcal{D}.\textsc{ReadBytecode}(a)$
\EndFor
\LineComment{Phase 3: Execute from fully-populated cache}
\For{$i \gets 1$ \textbf{to} $n$}
    \State Execute $T_i$ using $\mathit{cache}$ \Comment{All reads hit cache}
\EndFor
\end{algorithmic}
\end{algorithm}

\Cref{alg:ira-backup} summarizes the backup replay procedure.
The backup decompresses the hint, sorts all keys for sequential access (Phase~1), prefetches state via forward cursor traversal (Phase~2), and executes all transactions from the fully populated cache (Phase~3).

\subsection{Correctness and Safety}
\label{sec:correctness}

\parhead{Hint Completeness}
Since we generate hints by instrumenting the actual EVM execution on the primary, they capture exactly the set of keys that will be accessed during replay.
The deterministic nature of EVM execution guarantees that the backup, given the same block and state, will access the same keys.

\parhead{Advisory Hints}
The advisory property established in~\Cref{subsec:ira-architecture} manifests in \sysname{}-L through Zstandard's built-in integrity check on hint decompression and graceful fallback to reth's standard execution path on any coverage gap.
\Cref{sec:discussion} discusses adversarial-hint detection and mitigation. \Cref{subsec:threat-model} states the full adversary model.

\parhead{State Verification}
After executing each block, the primary computes a deterministic hash over all state changes which includes accounts, storage slots, and bytecodes, sorted by key for reproducibility.
The backup can optionally verify that its execution produced an identical hash, providing end-to-end correctness checking without requiring full Merkle state root computation.

\section{Evaluation}
\label{sec:eval}
\subsection{Configuration}
\label{sec:appendix:config}

\parhead{Hardware}
Apple M1 Pro ($10$-core: $8$ performance + $2$ efficiency), $32$ GB unified memory.
Benchmarks and trace databases on external $8$ TB NVMe SSD (Sabrent, USB enclosure).

\parhead{Software}
Rust 1.88 (edition 2024), Python 3.14 with matplotlib 3.10, numpy 2.4.
Apache Parquet (Arrow 54) with Zstandard compression for trace storage.
reth (commit 62abfdae) for trace collection (requires synced Ethereum mainnet node).

\parhead{Methodology}
We evaluate \sysname{}-L using the same two-week dataset of $\EvalBlockCount{}$ consecutive Ethereum mainnet blocks (blocks $24{,}019{,}447$--$24{,}120{,}246$). All benchmarks start with a cold OS page cache (purged before each run) to measure worst-case I/O performance. We compare four configuration families:
\begin{itemize}[nosep]
  \item \emph{Baseline}: unmodified reth executing blocks sequentially.
  \item \emph{Primary}: reth with hint generation enabled.
  \item \emph{Backup (sequential)}: backup replaying with hints using a single prefetch thread.
  \item \emph{Backup (parallel-$k$)}: backup replaying with $k$ concurrent prefetch threads ($k \in \{1, 2, 4, 6, 8, 10, 12, 16, 32, 64\}$). We report detailed per-block results for $k = 16$ and $k = 64$; the full sweep is used in the thread-scaling analysis (\Cref{sec:eval:thread-scaling}).
\end{itemize}

\subsection{Hint Overhead on Primary}
\label{sec:eval:primary-overhead}
The primary generates hints during normal block execution. We measure two components of overhead: \emph{hint construction} (collecting the access set) and \emph{hint serialization} (compressing and writing hints to disk). We compare against the unmodified baseline to quantify the total cost.

\parhead{Wall-Time Overhead}
\Cref{tab:primary-overhead} summarizes the overhead. The baseline executes all $\EvalBlockCount{}$ blocks in $\EvalBaselineWallTimeSec{}$ s, while the primary completes in $\EvalPrimaryWallTimeSec{}$ s---a $\EvalPrimaryWallOverheadPct{}\%$ increase. Of the hint-related work, construction (collecting the access set and querying the history index) accounts for $\EvalHintConstructTimeSec{}$ s ($\EvalHintConstructOverheadPct{}\%$ of aggregate execution time) and serialization (compressing and writing to disk) accounts for $\EvalHintWriteTimeSec{}$ s ($\EvalHintWriteOverheadPct{}\%$), totaling $\EvalHintTotalTimeSec{}$ s ($\EvalHintOverheadPct{}\%$). The remaining wall-time increase reflects secondary effects of instrumentation (e.g., memory pressure, cache interference).

\begin{table}[t]
\centering
\caption{Primary overhead breakdown. Wall time compares end-to-end execution; hint cost decomposes the overhead into construction and serialization as a fraction of aggregate execution time ($\EvalPrimaryExecTimeSec{}$\,s).}
\label{tab:primary-overhead}
\small
\begin{tabular}{lrr}
\toprule
\textbf{Component} & \textbf{Time (s)} & \textbf{\% of Exec Time} \\
\midrule
Baseline wall time       & $\EvalBaselineWallTimeSec{}$  & --- \\
Primary wall time        & $\EvalPrimaryWallTimeSec{}$   & --- \\
\quad Hint construction  & $\EvalHintConstructTimeSec{}$ & $\EvalHintConstructOverheadPct{}\%$ \\
\quad Hint serialization & $\EvalHintWriteTimeSec{}$     & $\EvalHintWriteOverheadPct{}\%$ \\
\quad \textbf{Hint total}& $\EvalHintTotalTimeSec{}$     & $\EvalHintOverheadPct{}\%$ \\
\midrule
Wall-time overhead       & \multicolumn{2}{c}{$+\EvalPrimaryWallOverheadPct{}\%$} \\
\bottomrule
\end{tabular}
\end{table}

\begin{figure}[!ht]
  \centering
  \includegraphics[width=0.9\columnwidth]{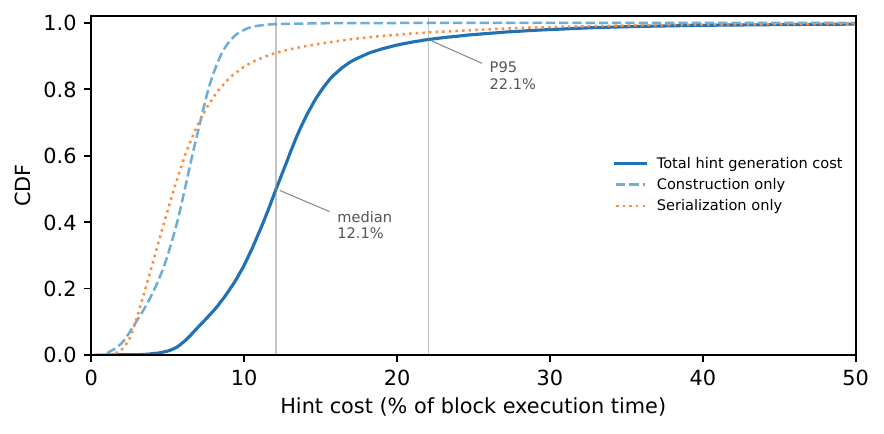}
  \caption{CDF of per-block hint cost as a percentage of block execution time. Each curve is an independent distribution: the total-cost curve is the CDF of per-block $(t_{\text{construct}} + t_{\text{serialize}}) / t_{\text{exec}}$, not the sum of the two component CDFs. Median total hint fraction is $\EvalHintFractionMedianPct{}\%$; P95 is $\EvalHintFractionPNinetyFivePct{}\%$.}
  \label{fig:per-block-hint-cost}
\end{figure}

\parhead{Per-Block Hint Cost}
\Cref{fig:per-block-hint-cost} shows the per-block distribution of hint cost as a fraction of execution time. The median per-block hint fraction is $\EvalHintFractionMedianPct{}\%$, with a $95$th-percentile of $\EvalHintFractionPNinetyFivePct{}\%$. Although construction accounts for more aggregate time than serialization (\Cref{tab:primary-overhead}), the serialization CDF has a heavier right tail: serialization time is roughly constant (${\sim}\EvalSerializationMedianMs{}$\,ms), so it dominates the per-block fraction for short-executing blocks, while construction scales with block size and its fraction remains bounded. Each CDF is sorted independently, so the total-cost curve is the distribution of per-block sums, not the pointwise sum of the two component curves.

In absolute time, construction spans a wide range (median $\EvalConstructMedianMs{}$\,ms, P95 $\EvalConstructPNinetyFiveMs{}$\,ms) and scales linearly with the number of unique storage keys ($r = \EvalConstructKeysCorr{}$, ${\sim}\EvalConstructPerKeyUs{}$\,$\mu$s/key). Serialization clusters tightly around ${\sim}\EvalWriteMedianMs{}$\,ms regardless of block size ($r = \EvalWriteKeysCorr{}$), dominated by the fixed cost of flushing the database write transaction. Construction therefore dominates aggregate cost (\Cref{tab:primary-overhead}) even though serialization dominates the per-block \emph{fraction} for short-executing blocks. The per-block hint fraction varies primarily because \emph{execution time} varies---driven by I/O and page cache state---not because hint cost itself is unpredictable.\iffullversion{} \Cref{sec:appendix:hint-cost} provides absolute-time CDFs, scaling plots, and a detailed variance decomposition.\fi

These per-block fractions (median $\EvalHintFractionMedianPct{}\%$, P95 $\EvalHintFractionPNinetyFivePct{}\%$) reflect a conservative, fully sequential measurement where hint construction and serialization run inline with block execution. In practice, both components are independent of execution correctness and can be overlapped with the next block's processing. Hint construction ($\EvalHintConstructOverheadPct{}\%$ of aggregate execution time; \Cref{tab:primary-overhead}) builds a map from keys already collected during EVM execution and queries the storage history index to annotate each key with its optimal read source\iffullversion{} (\Cref{sec:reth arch})\fi. Hint serialization ($\EvalHintWriteOverheadPct{}\%$) compresses and persists the hint to disk, requiring no executor state at all. Production reth already employs a multi-threaded architecture with a dedicated executor thread pool and a separate storage writer thread\iffullversion{} (\Cref{sec:reth arch})\fi; in this model, hint construction can execute on the existing thread pool and serialization can run on the storage thread, overlapping both with the next block's execution. Under such pipelining, the effective overhead approaches zero for all but the final block in a batch. Moreover, the cost is paid once by the primary and recovered on every backup: even with sequential prefetching, the backup replays $\EvalWallSpeedupSequential{}\times$ faster in wall time (\Cref{sec:eval:backup-speedup}), yielding a net system-wide speedup despite the primary's $\EvalPrimaryWallOverheadPct{}\%$ wall-time increase.

\begin{figure}[!ht]
  \centering
  \includegraphics[width=0.9\columnwidth]{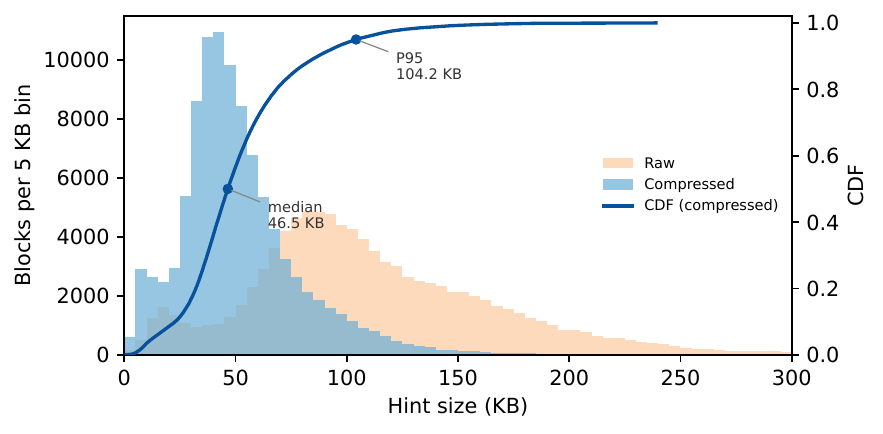}
  \caption{Distribution of per-block hint sizes. Histogram bins show block counts for raw (orange) and compressed (blue) sizes; the overlaid CDF (dark blue) marks the median ($\EvalHintCompressedMedianKB{}$ KB) and P95 ($\EvalHintCompressedPNinetyFiveKB{}$ KB) of compressed hints.}
  \label{fig:hint-size-cdf}
\end{figure}

\begin{figure}[!ht]
  \centering
  \includegraphics[width=0.9\columnwidth]{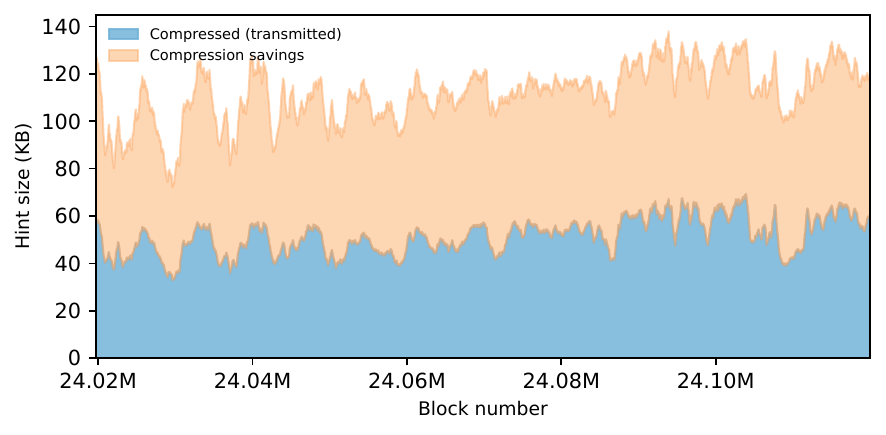}
  \caption{Hint sizes across blocks (500-block rolling average). The bottom layer shows the compressed size transmitted to backups; the top layer shows the bytes eliminated by Zstandard compression.}
  \label{fig:hint-size-per-block}
\end{figure}

\parhead{Hint Sizes}
\Cref{fig:hint-size-cdf} shows the distribution of per-block hint sizes. The median compressed hint is $\EvalHintCompressedMedianKB{}$ KB (mean $\EvalHintCompressedMeanKB{}$ KB), with a $95$th-percentile of $\EvalHintCompressedPNinetyFiveKB{}$ KB and a maximum of $\EvalHintCompressedMaxKB{}$ KB. Zstandard compression achieves a $\EvalHintCompressionRatio{}\times$ reduction from the raw representation (median $\EvalHintRawMedianKB{}$ KB raw). \Cref{fig:hint-size-per-block} confirms that these sizes remain stable across the entire $\EvalBlockCount{}$-block range, with no visible trend or drift.

A hint contains one fixed-size entry per unique storage key accessed in the block, encoding the key hash and its optimal read source. The raw hint size is therefore determined entirely by the key count: a block touching $\UniqueKeysMedian{}$ unique keys (the median; \Cref{sec:ethereum case study}) produces a hint of $\EvalHintRawMedianKB{}$ KB before compression. Zstandard is effective because, while the key hashes themselves are high-entropy, the accompanying metadata fields (source flags, offsets) are highly regular.

These hint sizes are negligible relative to the block data they accompany (${\sim}2$ MB per block): even the $95$th-percentile compressed hint adds only ${\sim}5\%$ to the block payload. On a $1$ Gbps link, transmitting the P95 hint takes under $1$ ms which is orders of magnitude less than the replay time saved. Hint size is also bounded by the EVM gas limit, which caps the number of storage operations per block. In practice, hints grow only when blocks are large, and large blocks are precisely those where replay savings are greatest (\Cref{sec:eval:backup-speedup}).

\subsection{Backup Speedup}
\label{sec:eval:backup-speedup}

We now evaluate how effectively hints accelerate backup replay. For each block, the backup reads the hint, prefetches the indicated keys into its MDBX cache, and then executes the block. We report the per-block speedup as $\text{speedup} = t_{\text{baseline}} / (t_{\text{wait}} + t_{\text{exec}})$, where $t_{\text{baseline}}$ is the baseline execution time, $t_{\text{wait}}$ is the time the executor stalls waiting for prefetch, and $t_{\text{exec}}$ is the backup's execution time.

\subsubsection{Per-Block Speedup Distribution}
\label{sec:eval:speedup-dist}

\begin{figure}[!ht]
  \centering
  \includegraphics[width=\columnwidth]{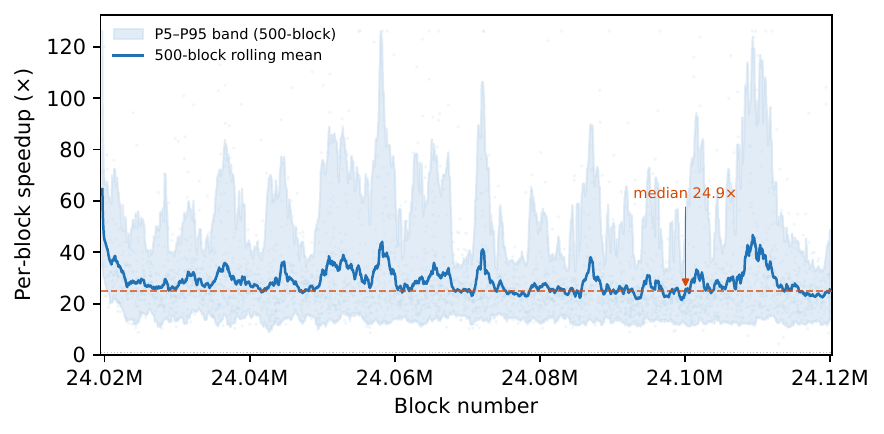}
  \caption{Per-block speedup (sequential backup vs.\ baseline) over the $\EvalBlockCount{}$-block trace. Each point is one block; the solid line is a $500$-block rolling mean and the shaded band spans the rolling P5--P95 within each window. The dashed line marks the overall median ($\EvalSpeedupSequentialMedian{}\times$).}
  \label{fig:speedup-per-block}
\end{figure}

\Cref{fig:speedup-per-block} shows per-block speedup across the entire block range for the sequential backup configuration;
\iffullversion
\Cref{tab:speedup-summary} in the appendix lists the key percentiles for all three configurations.
\fi

The backup achieves a median \emph{per-block} speedup of $\EvalSpeedupSequentialMedian{}\times$ (mean $\EvalSpeedupSequentialMean{}\times$) across the sequential configuration, with the parallel configurations yielding nearly identical distributions.
In terms of aggregate wall time, the sequential backup completes all $\EvalBlockCount{}$ blocks in $\EvalWallTimeSequentialSec{}$ s ($\EvalWallSpeedupSequential{}\times$ faster than the baseline's $\EvalBaselineTotalTimeSec{}$ s), parallel-$16$ in $\EvalWallTimeParallelSixteenSec{}$ s ($\EvalWallSpeedupParallelSixteen{}\times$), and parallel-$64$ in $\EvalWallTimeParallelSixtyFourSec{}$ s ($\EvalWallSpeedupParallelSixtyFour{}\times$). The gap between per-block and wall-time speedup for the sequential configuration arises because the $\EvalSlowerSequentialPct{}\%$ of blocks with regressions (\Cref{sec:eval:tail}) contribute disproportionately to aggregate wall time.

The speedup magnitude reflects the I/O-dominance of baseline execution (\Cref{sec:ethereum case study}): because storage I/O accounts for the majority of block execution time, prefetching the entire working set before execution begins eliminates nearly all of it, leaving only EVM computation.

\Cref{fig:speedup-per-block} reveals three regimes. First, an \emph{initial transient} in the first ${\sim}500$ blocks where speedup exceeds $50\times$: the baseline starts with a fully cold OS page cache, while the backup begins against an already-populated MDBX file and completes with zero wait time. As the baseline's page cache fills, its execution time drops and the speedup converges to steady state.

Second, the \emph{steady-state} regime (blocks $1{,}000$+) exhibits a stable rolling mean of $27$-$30\times$ with no meaningful drift and dividing the trace into $10$ equal segments yields segment means that vary by less than $\pm 10\%$ of the overall mean. This stability confirms that the speedup is determined by the per-block I/O fraction rather than by transient cache effects.

Third, \emph{periodic bursts} of $60$-$130\times$ appear throughout the trace. These high-speedup blocks have $2\times$ higher baseline execution time (more page cache misses) and $38\%$ more unique storage keys than the median, yet complete with zero backup wait time. The bursts are not sustained, the median cluster length is $1$ block. This indicates that they arise from individual blocks whose working sets happen to miss heavily in the baseline's page cache.

This $\EvalWallSpeedupSequential{}\times$ single-thread speedup exceeds the ${\sim}3\times$ that Amdahl's law predicts from the $\IoTimePct{}\%$ I/O share. That bound assumes the remaining $\ComputeTimePct{}\%$ is fixed compute. In practice, much of it is time spent servicing in-memory cache hits during execution, which \sysname{} changes fundamentally: it prefetches the working set into a flat, sorted vector, so these accesses also run faster through improved memory locality. \sysname{} accelerates both the I/O and the compute terms, so a bound derived from the I/O share alone does not apply.

\subsubsection{Tail Behavior}
\label{sec:eval:tail}

While the median speedup is substantial, we examine the fraction of blocks where the backup is \emph{slower} than the baseline\iffullversion{} (\Cref{tab:regression} in the appendix)\fi.

With sequential prefetching, $\EvalSlowerSequentialPct{}\%$ of blocks experience a regression\iffullversion{} (\Cref{tab:regression} in the appendix)\fi. The root cause is MDBX's use of memory-mapped I/O: each cold page access blocks the faulting thread until the kernel completes the read\iffullversion{} (\Cref{subsec:mdbx description})\fi, so a single prefetch thread issues at most one storage I/O at a time. When a block touches thousands of uncached B-tree pages, this serialization causes the executor to stall waiting for prefetch to finish, negating the I/O savings. Among these regressed blocks, the median speedup is $0.18\times$ ($5.5\times$ slower than baseline), with a mean of $0.20\times$ ($4.9\times$ slower). At $64$ prefetch threads, only $\EvalSlowerParallelSixtyFourPct{}\%$ of blocks remain slower than the baseline.

\subsubsection{Wait Time Analysis}
\label{sec:eval:wait-time}

\begin{figure}[!ht]
  \centering
  \includegraphics[width=0.9\columnwidth]{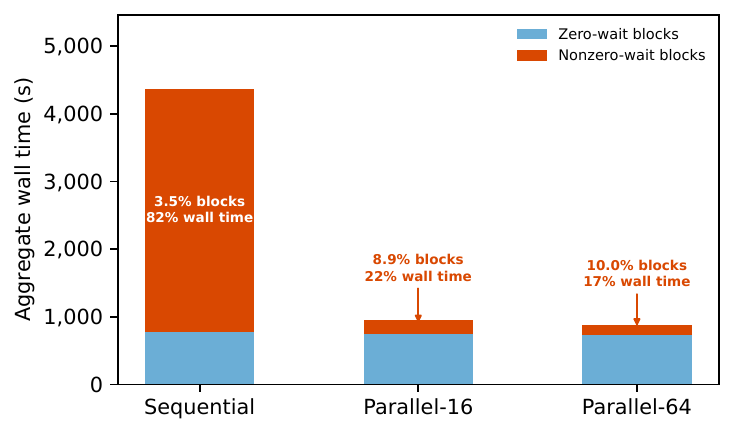}
  \caption{Aggregate wall time decomposed by executor wait. For each configuration, the bottom segment is wall time from blocks with zero wait; the top segment is from blocks where the executor stalled for prefetch.}
  \label{fig:wait-cdf}
\end{figure}

\Cref{fig:wait-cdf} decomposes aggregate wall time by whether the executor stalled waiting for prefetch to complete.

With sequential prefetching, $96.5\%$ of blocks complete with zero wait time, yet the remaining $3.5\%$ blocks, which are the same large-working-set blocks identified in \Cref{sec:eval:tail}, account for over $80\%$ of aggregate wall time. Parallel prefetching eliminates these stalls: with multiple threads, page faults are in flight simultaneously and the prefetch phase completes before the executor is ready to begin. This reduction in wait time is the primary driver of the wall-time improvement from sequential to parallel configurations, as execution time itself is unchanged.

\subsection{Thread Scaling}
\label{sec:eval:thread-scaling}

\begin{figure}[!ht]
  \centering
  \includegraphics[width=\columnwidth]{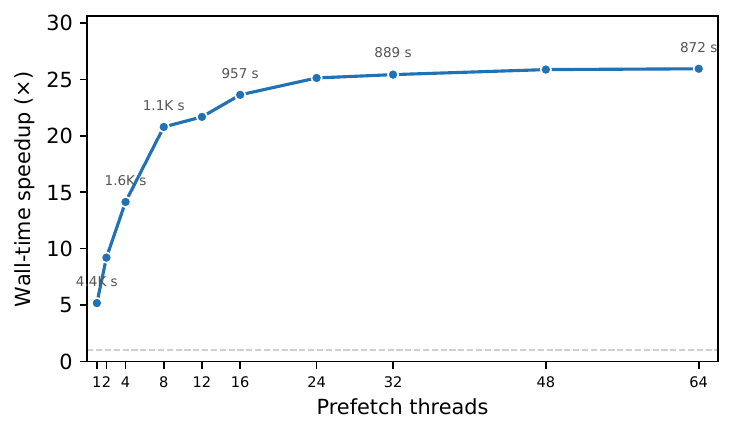}
  \caption{Wall-time speedup relative to the baseline as a function of prefetch thread count ($1$--$64$ threads). Labels show absolute wall time. The curve saturates around $8$--$16$ threads, consistent with I/O bandwidth limits of the NVMe SSD.}
  \label{fig:thread-scaling}
\end{figure}

\Cref{fig:thread-scaling} shows the wall-time speedup relative to the baseline as a function of prefetch thread count across ten configurations ($1$--$64$ threads). The curve exhibits three regimes. First, a \emph{steep scaling} region from $1$ to $8$ threads where wall time drops from $\EvalWallTimeSequentialSec{}$\,s to ${\sim}1{,}088$\,s ($4\times$ reduction), reflecting near-linear conversion of serial page faults into parallel I/O. Second, a \emph{diminishing-returns} region from $8$ to $16$ threads where wall time falls further to $\EvalWallTimeParallelSixteenSec{}$\,s ($\EvalScalingOneToSixteen{}\times$ over sequential) as the SSD's command queue approaches saturation. Third, a \emph{plateau} beyond $16$ threads: doubling from $16$ to $32$ yields only a $7\%$ reduction, and quadrupling to $64$ threads reaches $\EvalWallTimeParallelSixtyFourSec{}$\,s ($\EvalScalingSixteenToSixtyFour{}\times$ over $16$ threads), indicating that the I/O bandwidth of our NVMe SSD is fully saturated.

The steep initial gains reflect the conversion of serial page faults into parallel I/O: when a single thread prefetches sequentially, each MDBX page fault blocks until the SSD services the request. With multiple concurrent threads, page faults are in flight simultaneously, saturating the SSD's queue depth and achieving higher throughput. The knee at $8$-$16$ threads corresponds to the device's maximum queue depth being reached; additional threads merely compete for the same bandwidth. The optimal thread count is hardware-dependent since systems with higher-bandwidth storage (e.g., PCIe 5.0 NVMe, RAID arrays) may shift the knee to higher thread counts.

\subsection{Adversarial Hints}
\label{sec:eval:adversarial}

We measure the time for the latency of execution caused by a malformed hint. An adversary that sends a hint of the same size as the honest one but with non-existing keys, so accesses during execution miss the prefetched cache. In our system, this triggers the fallback to standard replay~(\Cref{sec:discussion}). We measure this cost with respect to the baseline cost of a system that nodes run today.

\italhead{Methodology}
We replay $\AdvHintBlockCount$ randomly sampled blocks against the reth state with a cold OS page cache, purged before each pass as in~\Cref{sec:appendix:config}. Each block is replayed two ways: \emph{baseline}, standard reth with no hints and on-demand reads; and \emph{malicious}, a parallel sorted-batch prefetch of the spurious hint followed by the fallback replay. The spurious hints match the per-block size of the captured hints (median $\AdvHintMedianKeys$ keys across the $\AdvHintCapturedBlocks$ captured blocks), and prefetching uses $\AdvHintPrefetchThreads$ threads. We report the per-block slowdown as the ratio of malicious to baseline replay time.

\italhead{Results}
Under the worst-case malicious hint, the backup replays at a median of $\AdvHintMaliciousMedianMs$\,ms per block against the baseline's $\AdvHintBaselineMedianMs$\,ms, a median slowdown of $\AdvHintSlowdownMedian\times$ (mean $\AdvHintSlowdownMean\times$). The tail stays bounded at $\AdvHintSlowdownPNinetyFive\times$ (P95) and $\AdvHintSlowdownPNinetyNine\times$ (P99). A malicious hint slows $\AdvHintSlowerPct\%$ of blocks relative to baseline, confirming that the adversary imposes a real but small cost rather than an unbounded one.

\italhead{Analysis}
The standard-replay fallback reads exactly the state the backup would have read without \sysname{}, so a malicious hint adds at most one wasted prefetch to an otherwise standard replay. This is akin to a peer relaying an incorrectly signed block or a block with incorrect metadata: the backup detects the invalid data during replay and discards the hint, falling back to the path it would have taken without \sysname{}. Standard peer-management mechanisms that drop or penalize peers on receiving invalid data limit the impact further, since a primary or relay that repeatedly delivers bad hints is disconnected. The slowdown therefore stays bounded and never affects correctness. We discuss detection and peer-level mitigation further in~\Cref{sec:discussion}.

\iffullversion
\subsection{Resource Usage}
\label{sec:eval:resource-usage}

A key property of \sysname{}-L is that hint-based prefetching does not require additional memory beyond what the baseline already uses. \Cref{tab:peak-rss} shows the peak resident set size (RSS) for each configuration.

\begin{table}[t]
\centering
\caption{Peak resident set size across configurations.}
\label{tab:peak-rss}
\small
\begin{tabular}{lr}
\toprule
\textbf{Configuration} & \textbf{Peak RSS (GB)} \\
\midrule
Baseline & $\EvalBaselineRssGB{}$ \\
Primary  & $\EvalPrimaryRssGB{}$ \\
Backup   & $\EvalBackupRssGB{}$ \\
\bottomrule
\end{tabular}
\end{table}

All three configurations consume comparable memory (${\sim}\EvalBaselineRssGB{}$ GB), dominated by the MDBX memory-mapped database. The primary's hint data structures (the access set) add negligible overhead: the set contains at most one entry per unique key in the current block (median $\UniqueKeysMedian{}$ keys), each requiring a few tens of bytes, totaling at most a few hundred kilobytes per block.

The backup's prefetch threads share the same memory-mapped file and do not allocate additional buffers. They simply trigger page faults that the OS services into the existing page cache. This design ensures that \sysname{}-L's memory footprint is determined by the database size, not by the hint mechanism.
\fi

\section{Related Work}
\label{sec:related}
\parhead{Optimal Cache Replacement}
Belady's MIN algorithm~\cite{beladyStudyReplacementAlgorithms1966}, proved optimal by Mattson et al.~\cite{mattsonEvaluationTechniquesStorage1970}, provides optimal offline cache replacement by evicting the item whose next access is furthest in the future. Some works use adaptive heuristics, e.g., the \emph{adaptive replacement cache}~\cite{megiddoARCSelfTuningLow2003} (ARC) and LIRS~\cite{jiangLIRSEfficientLow2002} which use heuristics to dynamically balance between recency and frequency, and LeCaR~\cite{vietriDrivingCacheReplacement2018} uses regret minimization to learn this balance online. Another approach is to use hardware prefetchers which predict future accesses to hide memory latency. Prefetch-aware replacement policies like PACMan~\cite{wuPACManPrefetchawareCache2011} and SHiP~\cite{wuSHiPSignaturebasedHit2011} adjust eviction priorities for prefetched blocks. A different approach is to approximate the MIN algorithm from history. The Hawkeye cache~\cite{jainBackFutureLeveraging2016} uses past access patterns to approximate MIN decisions. Jain and Lin~\cite{jainRethinkingBeladysAlgorithm2018} later observe that MIN minimizes \emph{total} misses but not \emph{demand} misses in the presence of prefetching, introducing Demand-MIN and Flex-MIN to navigate this tradeoff. Subsequent works like Mockingjay~\cite{shahEffectiveMimicryBeladys2022} and Glider~\cite{shiApplyingDeepLearning2019} use machine learning to improve prediction accuracy but require specialized hardware structures and still cannot exceed the accuracy ceiling imposed by predicting from past behavior alone. Recent deep learning approaches~\cite{shiHierarchicalNeuralModel2021,yangSGDPStreamGraphNeural2023,ganfureDeepPrefetcherDeepLearning2020} use neural networks to learn access patterns, achieving higher accuracy than traditional heuristics but remaining fundamentally limited to inference from history. Our work differs from these works in that we transmit exact future access information from primary to backup, enabling true optimal eviction without prediction.

\parhead{Hint-based systems}
Mowry et al.~\cite{mowryDesignEvaluationCompiler1992} presented the pioneering work with the idea of using compiler based automation to optimize prefetching. The TIP system~\cite{pattersonInformedPrefetchingCaching1995} by Patterson et al. was one of the early works proposing the \emph{hint}-based paradigm where the applications advise the OS with \emph{advisory hints} to improve prefetching and caching optimizations. HintStor~\cite{geHintStorFrameworkStudy2022} presents a framework to propagate I/O hints across the storage stack in heterogeneous storage systems. Mandal et al.~\cite{mandalUsingHintsImprove2016} demonstrate hints at the block layer of storage for deduplication, where applications mark unique writes (\texttt{NODEDUP}) and prefetch requests (\texttt{PREFETCH}) to avoid redundant hash computations. Modern OS provide system calls such as \texttt{madvise}~\cite{Madvise2LinuxManual} and \texttt{fadvise}~\cite{Posix_fadvise2LinuxManual} where applications can advise the OS on the access pattern such as \texttt{random}, \texttt{sequential}, \texttt{will-need}, and \texttt{don't need}. Database systems such as Oracle's \texttt{INDEX} and \texttt{USE\_NL} hints allow developers to override the query planner run before executing a query. These approaches share the limitation that hints are \emph{manually} specified and require domain knowledge, or \emph{statically derived} which are limited to predictable patterns like loops. In contrast, our framework \sysname{} generates hints automatically from the primary's execution trace, providing complete and exact future access information without programmer effort.

\parhead{Replay in Replicated Databases}
Modern databases build on ARIES-style write-ahead logging~\cite{mohanARIESTransactionRecovery1992}, which established the three-phase recovery protocol (analysis, redo, undo) now used across the industry~\cite{mohanRepeatingHistoryARIES1999}. In primary-backup replication, backups continuously replay the primary's log to maintain synchronized state. Deterministic databases such as Calvin~\cite{thomsonCalvinFastDistributed2012} and Aria~\cite{luAriaFastPractical2020} ensure replicas converge to identical states by replaying transaction logs in a deterministic order without coordination. A key challenge is the \emph{parallelism gap}: primaries execute transactions concurrently, but backups traditionally replay logs serially. KuaFu~\cite{chuntaohongKuaFuClosingParallelism2013} addresses this by tracking dependencies to enable parallel replay, demonstrating that serial replay sustains less than one-third of primary throughput thus underscoring the need for systems such as \sysname{}. Cloud-native databases have further evolved replay architectures: Amazon Aurora~\cite{verbitskiAmazonAuroraDesign2017} pushes redo log application to a distributed storage tier, treating the log as the database itself, while Socrates~\cite{antonopoulosSocratesNewSQL2019} (Azure SQL Hyperscale) uses a four-tier architecture with dedicated Page Servers for log application. PolarDB-SCC~\cite{yangPolarDBSCCCloudNativeDatabase2023} tackles the consistency challenges arising from asynchronous log propagation to read-only replicas. While these works focus on replay correctness, parallelism, or architectural placement of log application, none optimize \emph{cache eviction decisions} during replay. \sysname{} is complementary: it accelerates replay within any of these architectures by exploiting the primary's knowledge of future accesses to achieve Belady-optimal caching.

\parhead{Blockchain State Synchronization}
Blockchain nodes face significant replay challenges during both initial synchronization and ongoing block validation. Modern Ethereum clients use staged sync architectures~\cite{ErigonEthStagedsync,ParadigmxyzReth2026} that pipeline header, body, and execution stages; despite these optimizations, the execution stage remains the primary bottleneck. To enable parallel validation, some systems require transactions to declare data dependencies upfront: Ethereum's EIP-2930~\cite{proposalsEIP2930OptionalAccess} introduced optional access lists, though adoption remains low~\cite{heimbachDissectingEIP2930Optional2023}; Solana's Sealevel runtime~\cite{yakovenkoSolanaNewArchitecture2025} mandates read/write set declarations. For systems without mandatory declarations, Block-STM~\cite{gelashviliBlockSTMScalingBlockchain2022} provides optimistic parallel execution, and Forerunner~\cite{chenForerunnerConstraintbasedSpeculative2021} exploits speculative pre-execution. While these works optimize sync architecture or transaction scheduling, none address \emph{cache eviction} during replay. Unlike optional, user-declared access lists, \sysname{}'s hints are execution-derived, complete by construction, and carry a per-key source annotation~(\Cref{sec:hint-generation}) that eliminates history-index seeks, enabling optimal caching during the execution stage. \Cref{sec:appendix:extended-related} provides extended discussion.

\parhead{Speculative and parallel execution}
Parallel transaction execution has been extensively studied. Kung and Robinson's optimistic concurrency control (OCC)~\cite{kungOptimisticMethodsConcurrency1981} established the read-validate-write paradigm; Silo~\cite{tuSpeedyTransactionsMulticore2013} demonstrated OCC can achieve nearly $700{,}000$ transactions per second on multicore hardware. For high-contention workloads, Doppel~\cite{narulaPhaseReconciliationContended2014} introduced phase reconciliation, and deterministic approaches like BOHM~\cite{faleiroRethinkingSerializableMultiversion2015} pre-compute dependencies to enable parallel execution. \sysname{} is orthogonal to these approaches: even when transactions execute in parallel, each thread must manage its cache. By providing exact future access information, \sysname{} enables optimal caching \emph{within} each parallel execution context. \Cref{sec:appendix:extended-related} provides extended discussion.

\section{Conclusion}
\label{sec:conclusion}
We presented \sysname{}, a general framework for accelerating backup replay in primary-backup replication by transmitting compact hints that encode the primary's knowledge of future access patterns. The framework applies to any storage-bound, deterministic primary-backup system including replicated databases, distributed key-value stores, and blockchain platforms. We demonstrated its practicality through \sysname{}-L, a concrete instantiation for Ethereum block synchronization implemented in the production client reth, showing significant replay speedups at low hint overheads.

\section*{Disclaimer}

Case studies, comparisons, statistics, research and recommendations are
provided ``AS IS'' and intended for informational purposes only and should not
be relied upon for operational, marketing, legal, technical, tax, financial or
other advice. Visa Inc.\ neither makes any warranty or representation as to the
completeness or accuracy of the information within this document, nor assumes
any liability or responsibility that may result from reliance on such
information. The Information contained herein is not intended as investment or
legal advice, and readers are encouraged to seek the advice of a competent
professional where such advice is required.

These materials and best practice recommendations are provided for
informational purposes only and should not be relied upon for marketing, legal,
regulatory or other advice. Recommended marketing materials should be
independently evaluated in light of your specific business needs and any
applicable laws and regulations. Visa is not responsible for your use of the
marketing materials, best practice recommendations, or other information,
including errors of any kind, contained in this document.

\bibliographystyle{plainurl}
\bibliography{extra-references.bib,Ira-bibliography-zotero}

\appendix

\section{Ethical Considerations}

\italhead{Scope}
\sysname{}-L's main contribution is a performance optimization for honest Ethereum block replay, accompanied by a security analysis of its trade-offs~(\Cref{subsec:threat-model,sec:eval:adversarial}). Hints are advisory and independently verified by the backup~(\Cref{sec:correctness,subsec:threat-model}): a primary capable of proposing a malicious block before \sysname{}-L has the same capability after, and a backup that ignores hints reproduces the unmodified baseline. We structure the analysis below using the Menlo principles~\cite{kenneallyMenloReportEthical2012}.

\italhead{Stakeholders}
Four groups are affected:
(i)~Ethereum node operators (validators, RPC providers, archive nodes);
(ii)~block proposers that emit hints;
(iii)~Ethereum users whose transactions are validated; and
(iv)~the broader ecosystem (protocol developers, client implementers, researchers).

\italhead{Respect for Persons}
Experiments use only public mainnet data collected on our own archive node. No private data, no human subjects, no identifying information about validator-running entities. Transactors are pseudonymous, and the storage-access patterns our hints encode are revealed on-chain after confirmation, so the hint channel adds no new information about identifiable persons. Adoption is opt-in: no stakeholder must interact with \sysname{}-L to remain a valid participant.

\italhead{Beneficence}
We weigh benefits against bounded harms per stakeholder.
\begin{itemize}[nosep]
  \item \emph{Operators} gain a $\EvalWallSpeedupParallelSixteen{}\times$ backup speedup (\Cref{sec:eval:backup-speedup}), lowering the commodity-hardware bar for running a full or archive node. Adversarial hints cause bounded slowdown, mitigated by standard-replay fallback (\Cref{sec:correctness}); an adopting node is never worse off than a non-adopting one. We rejected authoritative (non-advisory) hints because a Byzantine proposer could then corrupt backup state.
  \item \emph{Proposers} pay $\EvalHintOverheadPct{}\%$ in hint-generation overhead (\Cref{sec:eval:primary-overhead}) and gain no MEV, fee, or ordering capability. Malformed hints waste backup CPU but are bounded and discarded at the gossip layer, analogous to invalid block headers or bodies.
  \item \emph{Users} benefit from a tighter attestation window and a weakened verifier's dilemma~\cite{luuDemystifyingIncentivesConsensus2015}, with no direct harm and no new privacy surface.
  \item \emph{Ecosystem}: the client-agnostic hint format (\Cref{sec:hint-generation}) and open reference implementation make the benefits portable. An asymmetric rollout could widen the short-run cost gap between hint-aware and unmodified nodes, mitigated by releasing the implementation alongside the paper.
\end{itemize}

\italhead{Negative applications considered}
We flag two:
\begin{itemize}[nosep]
  \item \emph{Cheaper on-chain surveillance}: a lower archive-node cost also lowers the cost of large-scale address-activity indexing for actors hostile to transactor privacy. On balance, well-resourced surveillance actors already have this capability, and the decentralization gain for honest operators outweighs the marginal reduction.
  \item \emph{Hint-gossip side channel}: timing the arrival of hints could in principle leak storage-access sets earlier than today's block gossip, whose propagation is already an observable channel studied for validator deanonymization and topology inference~\cite{heimbachDeanonymizingEthereumValidators2024,zhaoDEthnaAccurateEthereum2024}. The signal is weak (hints come only from the primary, and block contents are revealed within seconds regardless).
\end{itemize}

\italhead{Justice}
Benefits are not restricted to any operator class. The design is client-agnostic, the implementation is openly licensed, and the hint format has no hardware, OS, or ISA dependency (\Cref{sec:hint-generation}). Evaluation singles out no individual or organization; all statistics are aggregated over the $\EvalBlockCount{}$ public mainnet blocks studied (\Cref{sec:ethereum case study}).

\italhead{Respect for Law and Public Interest}
We use only public mainnet data and circumvent no protocol rule, client license, or applicable regulation. No vulnerability disclosure is triggered. The public-interest case is the decentralization and consensus-security gain. The artifact accompanies the paper~\cite{bhatIraEfficientTransaction2026a}.

\italhead{Decision}
Under Beneficence, the benefits (lower replication cost, stronger consensus-layer security through a weakened verifier's dilemma, broadly distributed access) exceed the bounded and mitigatable harms (adversarial-hint slowdown, short-run rollout gap, marginal surveillance cost). A Respect-for-Persons analysis reaches the same conclusion: interaction is only via public on-chain data, no consent violation, no new Byzantine capability granted to the primary. We determine that publication is appropriate.

\section{Open Science}

We are committed to reproducibility. The following artifacts are available:

\parhead{Available Artifacts}
\begin{itemize}[nosep]
  \item \emph{Source code}: Our modifications to reth implementing hint generation (primary) and hint-based prefetching (backup), along with trace collection instrumentation. Available at~\cite{bhatAdithyabhatkajakeIrars2026}.
  \item \emph{Analysis scripts}: Python scripts for computing all statistics and generating all figures in this paper.
  \item \emph{Benchmark results}: CSV files containing raw timing data for all experiments (hint overhead, backup speedup, thread scaling).
  \item \emph{Trace statistics}: Aggregated statistics from our workload analysis (operation counts, key distributions, per-block metrics) in CSV format.
\end{itemize}
The data artifacts (pre-generated hints, Parquet traces, and state hashes) are archived at \url{https://doi.org/10.5281/zenodo.20262361}.

\parhead{Artifacts Not Shared}
\begin{itemize}[nosep]
  \item \emph{Ethereum archival node}: Reproducing trace collection requires a fully synced Ethereum archival node (${\sim}3$ TB). We cannot redistribute this data due to its size, but it is freely available to anyone running an archival node.
  \item \emph{Raw execution traces}: The Parquet trace files (${\sim}12$ GB) are derived from publicly available blockchain data. We provide aggregated statistics and the collection scripts; researchers with archival access can regenerate the traces.
\end{itemize}

\parhead{Reproducibility}
The repository includes a \texttt{README.md} with instructions for: (1)~building the modified reth client, (2)~running benchmarks using pre-generated hints, and (3)~regenerating all figures from the provided CSV data. Researchers without archival node access can reproduce all evaluation results using our pre-computed hint database and benchmark CSVs.

\section{Generalization to a Second Temporal Window}
\label{sec:appendix:second-window}

To check that \sysname{}-L's speedup generalizes, we re-ran the full pipeline of~\Cref{sec:eval} (identical binaries, hardware, and cold-cache protocol) on blocks $20,850,000$--$20,950,000$ ($\SecondWindowBlockCount{}$ blocks, September $28$ to October $12$, $2024$, $14$ months earlier). The median per-block speedup is $\SecondWindowSpeedupParallelSixteenMedian{}\times$ (parallel-$16$), aggregate wall-clock speedup is $\SecondWindowWallSpeedupParallelSixteen{}\times$, primary overhead is $\SecondWindowHintOverheadPct{}\%$, and only $\SecondWindowSlowerParallelSixteenPct{}\%$ of blocks regress. These are uniformly lower than the December $2025$ headline ($\EvalSpeedupParallelSixteenMedian{}\times$, $\EvalWallSpeedupParallelSixteen{}\times$, $\EvalHintOverheadPct{}\%$), and the gap is a workload effect: re-running the trace pipeline of~\Cref{sec:ethereum case study} on the October window yields a median of $\SecondWindowStorageOpsMedian{}$ storage operations per block ($\SecondWindowStorageOpsRatio{}\times$ below December's $\StorageOpsMedian{}$) and $\SecondWindowUniqueKeysMedian{}$ unique storage keys ($\SecondWindowUniqueKeysRatio{}\times$ below December's $\UniqueKeysMedian{}$), yet median baseline execution time is $\SecondWindowBaselineMsRatio{}\times$ lower ($\SecondWindowBaselineMedianMs{}$ versus $\BaselineMedianMs{}$ ms). Protocol-priced work scaled $\SecondWindowStorageOpsRatio{}\times$ while wall-clock I/O cost scaled $\SecondWindowBaselineMsRatio{}\times$, reflecting state-trie growth that lengthens cold-page lookups; \sysname{}-L's amplification tracks this I/O fraction, so the heavier later baseline yields the larger speedup. The same mechanism delivers an order-of-magnitude acceleration in either window.

\section{Discussion}
\label{sec:discussion}
\parhead{Application scope} Our framework has applications beyond Ethereum. The key requirements for \sysname{} to be applicable in a distributed database system are: (i) primary-backup architecture, i.e., there exists a node that executes first and the others follow, (ii) replay is storage-bound, where I/O is a bottleneck, and (iii) the execution is deterministic. Systems such as the following potentially benefit from our solution:
\begin{enumerate}
  \item \emph{Replicated databases}, e.g., MySQL Group Replication~\cite{MySQLMySQL802025}, MariaDB Galera~\cite{CertificationBasedReplicationGalera2025}, PostgreSQL streaming replication~\cite{262LogShippingStandby2025}, CockroachDB~\cite{taftCockroachDBResilientGeoDistributed2020}, TiDB~\cite{huangTiDBRaftbasedHTAP2020}, and deterministic databases like Calvin~\cite{thomsonCalvinFastDistributed2012}, use primary-backup architecture to deterministically replay database operations. All of them have nodes with complete execution traces that can assist replicas during replication. 
  \item \emph{Distributed key-value stores}, e.g., FoundationDB~\cite{zhouFoundationDBDistributedUnbundled2021}, use consensus-based replication where followers replay the primary's logs. For stores with large state that exceeds memory, hints can naturally encode access patterns to improve performance.
  \item \emph{Blockchain systems} such as Solana~\cite{yakovenkoSolanaNewArchitecture2025}, Sui~\cite{SuiSmartContracts2025}, Aptos~\cite{AptosBlockchainSafe2022,gelashviliBlockSTMScalingBlockchain2022}, Cosmos~\cite{buchmanLatestGossipBFT2019,CosmosStackDeveloper2025}, and the Hyperledger family~\cite{androulakiHyperledgerFabricDistributed2018} have validators that propose blocks while other nodes replay using a VM layer, and are strong candidates for \sysname{}'s hint-based acceleration. Layer-$2$ rollups (sequencer as primary) and sharded chains (per-shard primary-backup replay) are natural fits. \sysname{} needs no single leader either: in leaderless DAG-based protocols such as Narwhal~\cite{danezisNarwhalTuskDAGbased2021} and Bullshark~\cite{spiegelmanBullsharkDAGBFT2022}, any validator that has executed a batch can gossip its hints as an advisory byproduct, and access sets from different proposers are unionable.
\end{enumerate}

\parhead{Deployment considerations} 
\italhead{Layer-1 optional}
Hint correctness need not be enforced by the underlying synchronization protocol, whether in database replication or in blockchain block proposal.

\italhead{Incentive Mechanisms} We can incentivize our system with rewards. Today, Ethereum's EIP-2930 access lists~\cite{proposalsEIP2930OptionalAccess} incentivize transactions that declare an access list (a list of accounts accessed by the transaction) with discounted gas costs, though adoption remains low~\cite{heimbachDissectingEIP2930Optional2023}. If the access list is wrong, the user pays more gas. A similar mechanism can ensure that block producers generate correct hints.

\italhead{Security considerations} Our hint generation does not rely on the primary being non-faulty. On omission of hints, nodes can simply revert to a backup caching mechanism such as LRU. However, systems operating under stronger trust models, e.g., those using a Trusted Execution Environment (TEE) or crash-fault models, can take advantage of the non-faulty primary and provide additional hints that can further speed up verification as well as replay. Note that hints reveal access patterns, which may be a privacy consideration in permissioned blockchains where transaction contents are confidential.

\italhead{Security implications} Faster backup replay carries two security consequences. First, it narrows the verifier's dilemma~\cite{luuDemystifyingIncentivesConsensus2015}: cheaper re-execution raises a rational validator's incentive to verify each block, making honest validation the dominant strategy. Second, it improves resilience to denial-of-service: slow state access is a known attack surface, whether through mispriced execution~\cite{perezBrokenMetreAttacking2020} or deliberate state-storage bloat~\cite{heNurgleExacerbatingResource2024}, and \sysname{}-L's lower replay latency shrinks the window such attacks exploit. Both consequences follow from the optimization and remain orthogonal to its safety guarantees.

\italhead{Corrupted hints}
A peer delivering a corrupted or malicious hint is analogous to a peer delivering a corrupt block: in both cases, the recipient detects the fault and discards the invalid data. For hints, detection is straightforward: decompression failure (Zstandard integrity check), a malformed entry count, or a cache miss during execution all cause the backup to discard the hint and fall back to standard replay. Because hints are advisory and never affect state correctness, the impact of a corrupted hint is bounded to a single block's replay latency. Two further mechanisms bound this cost: the backup can heuristically upper-bound the number of storage accesses from the block's gas and cap the accepted hint size accordingly, rejecting oversized hints before prefetching; and prefetching hints in sorted batches confines I/O and cache pollution to that per-block working set, so even a maximally-spurious same-size hint cannot evict state beyond a single block's footprint. Our measurements confirm this bounds the slowdown to a small factor over standard replay~(\Cref{sec:eval:adversarial}).

We treat the problem of \emph{identifying and penalizing} peers that deliver corrupt hints as orthogonal to our work. Ethereum's networking stack already provides mechanisms for this: the consensus layer's gossipsub peer scoring~\cite{SpecsPubsubGossipsub} tracks peer behavior and disconnects peers whose scores fall below a threshold, while the execution layer's devp2p protocol supports explicit disconnect codes for protocol violations~\cite{EthereumDevp2p2026}. These same mechanisms can be extended to penalize peers that consistently deliver invalid hints, requiring no changes to \sysname{} itself.

\parhead{Limitations and future work}
Our implementation targets a single-threaded execution model. Concurrent execution would require coordination for hint data structures, potentially through partitioning hints by key ranges, usage of synchronization primitives and barriers. Notably, \sysname{} can also assist concurrent execution with hints that aid in improving concurrent replay. We can extend the hints to support parallel replay across multiple threads and explore partitioned hint structures to avoid synchronization overheads. Another limitation of our work is that it assumes the replay workload is storage-bound. Compute-bound replay, such as zk-proving or crypto-heavy opcodes, is out of scope and left to future work. Beyond cache eviction, hints could guide write batching, memory allocation, and as already stated, concurrency to further accelerate replay.

\iffullversion
\section{Reth Architecture}%
\label{sec:reth arch}

Ethereum uses two clients: a consensus client that follows a beacon chain to determine the blockchain that is valid, and an execution client that executes the blocks from the consensus client. They work together to keep a node in sync, validate incoming blocks, and to propose new blocks. This section describes the architecture of Reth, a Rust-based high performance execution client, in particular, we focus on the block validation and execution architecture relevant to understanding our hint-based optimization. \Cref{fig:reth arch block production validation} presents a schematic summary of the architecture that we are interested in. 

\subsection{System components}

The execution client consists of four primary components that communicate through message-passing channels:
\begin{enumerate}
  \item \parhead{Engine handler} This is the central coordinator that receives block validation requests from the consensus layer and orchestrates the validation pipeline. It runs in a dedicated thread and processes requests sequentially to ensure consistency without complex synchronization. It contains two sub-components:
        \begin{itemize}[nosep]
          \item \italhead{In-memory chain} It maintains an in-memory tree of recently validated blocks indexed by block hashes. Blocks remain here until they are finalized, allowing the engine for quick-access to the current chain. It also buffers blocks with unknown parents.
          \item \italhead{Validator} It validates blocks against consensus rules (format, limits, conditions) before and after the execution. 
        \end{itemize}
  \item \parhead{Executor} Executes transactions through the EVM and computes the post-execution state. The executor runs on a small thread pool ($3$ threads), and also performs operations such as transaction execution, and signature recovery from the transaction signers. It maintains a block-level cache for state access and supports speculative pre-warming by executing transactions in parallel to populate the cache before sequential execution.  
  \item \parhead{Block builder} Constructs new blocks when the node is selected as proposer. The builder selects transactions from the mempool, invokes the executor to verify validity and compute gas, assembles the final block.
  \item \parhead{Storage layer} Handles durable storage of finalized blocks. Runs on a dedicated thread and receives block batches from the engine handler. 
\end{enumerate}

\begin{figure}[!ht]
  \centering
  \includegraphics[width=\linewidth]{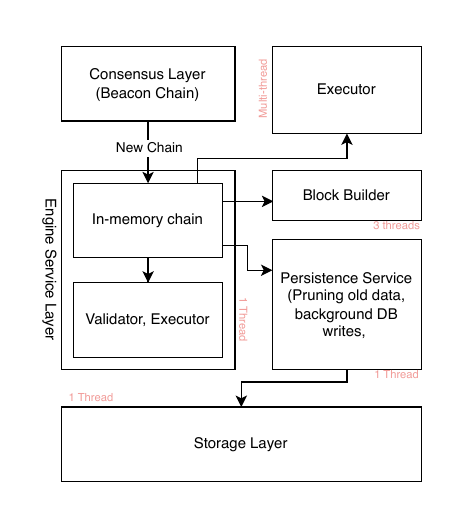}
  \caption{Reth architecture for block production and validation}\label{fig:reth arch block production validation}
\end{figure}

\parhead{State access and caching} The executor maintains a large in-memory cache (default $9$ GB) partitioned by data type:
\begin{itemize}[nosep]
  \item \italhead{Storage cache ($\sim 8$ GB, $89\%$)}: maps address-slot pairs to storage values. This is justified since it is the most dominant state access in smart contract execution.
  \item \italhead{Account cache ($\sim 500$ MB, $5.5\%$)} maps addresses to account metadata (balance, nonce, code hash). While these are frequently accessed, the data per account is small.
  \item \italhead{Bytecode cache ($\sim 500$ MB, $5.5\%$)} maps code hashes to contract bytecode. The values in these cache are immutable, but individual contracts can be large, e.g., $24$ KB.
\end{itemize}

The cache uses time-based eviction with a $2$-hour time-to-live and $1$-hour idle timeout. Entries are weighted by their actual memory footprint rather than count, ensuring the memory budget is respected regardless of entry size.

Cache misses fall through to MDBX. Historical data (old blocks, transactions, receipts) is stored in separate immutable files to reduce database size and improve locality for recent state.

\subsection{Storage}%
\label{subsec:reth-storage}

In this section, we focus on the storage layers relevant to the state access for block execution. Reth's storage layer uses three categories of tables to serve both current and historical state queries.

\parhead{Plain State Tables}
These store the current state of the chain. These tables are updated as new blocks are executed and represent the most recent state.

\italhead{Plain Account State Tables} These map each address to its current account (balance, nonce, code hash).

\italhead{Plain Storage State} These map each (address, slot) pair to its current storage value. 

\parhead{Change Set Tables}
These record what changed in each block, storing the value before the change. Together, change sets enable \emph{time travel}, i.e., to recover the state at any historical block, one can start from the current plain state and walk backwards through change sets.

\italhead{Account Change Sets} These are keyed by block number and stores the previous account state for every account modified in that block.

\italhead{Storage Change Sets} is keyed by (block number, address) and stores the previous storage value for every slot modified in that block.

\parhead{History Index Tables}
These provide efficient lookups to find which blocks modified a given key, avoiding a linear scan through change sets. The bitmaps use roaring bitmap compression and are sharded into chunks of 2,000 entries for bounded per-entry size.

\italhead{Accounts History} These map each address to a compressed bitmap of block numbers where that address was modified.

\italhead{Storages History} maps each (address, slot) pair to a compressed bitmap of block numbers where that slot was modified.

\subsection{MDBX database}%
\label{subsec:mdbx description}

All tables described above are stored in MDBX~\cite{LibmdbxOneFastest}, an embedded key-value database derived from LMDB. MDBX organizes data in B$^+$-trees and uses memory-mapped I/O to expose database pages directly in the process address space. Rather than issuing explicit read system calls, Reth accesses database pages by dereferencing pointers into the mapped region; the operating system's virtual memory subsystem transparently faults in pages from disk on demand.

\parhead{Implications for performance}
This design has two consequences that are central to our optimization. First, the cost of a database read depends on whether the accessed page resides in the OS page cache. A warm read (page already cached) completes with a page-table walk and a memory load, while a cold read triggers a page fault that blocks the calling thread until the page is read from storage. Because MDBX does not maintain its own buffer pool, the OS page replacement policy governs the eviction, which has no knowledge of the application's access pattern. Second, each historical state lookup traverses a B+ tree in the history index table, touching $O(\log n)$ pages, any of which may trigger a page fault. Under memory pressure or when the working set exceeds available RAM, these random traversals incur multiple storage round-trips per lookup. Our system \sysname{}-L tackles this bottleneck by supplying source hints that eliminate redundant index traversals.

\parhead{Why multi-threaded prefetching is necessary}
Although \sysname{}'s hints identify exactly which keys will be accessed, prefetching them efficiently is complicated by MDBX's use of memory-mapped I/O. Because reads are pointer dereferences into the mapped region rather than explicit system calls, a cold page access manifests as a synchronous page fault that blocks the faulting thread until the kernel completes the I/O. There is no mechanism to issue an asynchronous read against an \texttt{mmap}'d region from user space. While \texttt{madvise(MADV\_WILLNEED)}~\cite{Madvise2LinuxManual} can hint to the kernel that pages will be needed soon, it is purely advisory: the kernel may defer or ignore the request, and critically, it provides no completion notification, so the application cannot determine when pages are actually resident. A single-threaded prefetcher would therefore serialize all page faults, issuing at most one storage read at a time and failing to exploit the parallelism available in modern NVMe devices (which support tens of thousands of concurrent I/O operations). To overcome this, \sysname{}-L dispatches prefetch requests across a pool of threads: each thread touches a different B$^+$-tree path, absorbs its own page fault independently, and the OS services the resulting I/O requests concurrently through the NVMe submission queue. This helps reduce the impact of our inability to prefetch data.

\subsection{Reth Baseline Architecture}%
\label{subsec:reth baseline arch}

We build a benchmark tool that replays historical blocks to measure execution performance. It simulates production behavior for block validators.

\parhead{Components}
\begin{itemize}[nosep]
  \item \italhead{Block Processor} Iterates through blocks sequentially, coordinating the per-block cache and executor for each block. Records execution time and memory usage.
  \item \italhead{Per-Block Cache} A temporary cache created fresh for each block. During execution, it accumulates all state accessed by transactions (accounts, storage slots, bytecode). The executor reads and writes state through this cache. After the block completes, its contents are merged into the cross-block cache.
  \item \italhead{Executor} Runs transactions sequentially through the EVM. Reads state from the per-block cache and writes resulting state changes back to it.
  \item \italhead{Cross-Block Cache} A persistent LRU cache that survives across blocks. Contains:
        \begin{itemize}[nosep]
          \item Accounts ($100$K entries)
          \item Storage slots ($1$M entries)
          \item Bytecode ($10$K entries) 
        \end{itemize}
  When the per-block cache encounters a miss, it queries the cross-block cache. After each block, touched state is merged back, so block $N+1$ benefits from state accessed during block $N$.
  \item \italhead{Storage} The underlying database (reth's MDBX) opened in read-only mode. The cross-block cache queries storage on cache misses to fetch historical state.
\end{itemize}

\parhead{Data Flow}
\begin{enumerate}
  \item The block processor initializes a fresh per-block cache and invokes the executor
  \item The executor reads and writes state through the per-block cache
  \item On cache miss, the per-block cache queries the cross-block cache
  \item On cache miss, the cross-block cache queries storage
  \item After execution completes, the per-block cache contents are merged into the cross-block cache
  \item The process repeats for the next block
\end{enumerate}

\fi

\iffullversion
\section{\sysname{}-L Architecture}

In this section, we provide details of \sysname{}-L implementation in Rust.

\subsection{Reth Primary}%

\emph{reth-primary} is the hint generator that simulates a block proposer. It replays blocks, collects all state keys accessed during execution, and writes hints that enable other backups to prefetch state for faster replay. \Cref{fig:iral primary architecture} presents an illustration of our architecture.
\parhead{Components}
\begin{enumerate}
  \item \italhead{Block Processor} Iterates through blocks sequentially, coordinating execution, hint construction, and output writing.
  \item \italhead{Key Collector} An EVM inspector that records accessed keys by intercepting opcodes:
        \begin{itemize}[nosep]
          \item \texttt{SLOAD}/\texttt{SSTORE}: storage slot accesses                                                                  
          \item \texttt{BALANCE}/\texttt{SELFBALANCE}: account accesses
          \item \texttt{CALL}/\texttt{STATICCALL}/\texttt{DELEGATECALL}: bytecode accesses                                                      
          \item \texttt{EXTCODECOPY}/\texttt{EXTCODESIZE}: bytecode/account accesses
        \end{itemize}
        Additionally, we also add transaction senders, recipients, and the block beneficiary (coinbase) to account hints.
  \item \italhead{Executor} Runs transactions through the EVM with the inspector from the Key Collector attached. The inspector intercepts EVM opcodes to capture all state accesses without modifying execution behavior.
  \item \italhead{Source Tracker} For each storage key, queries the storage history index to determine the optimal read source:
        \begin{itemize}[nosep]
          \item  \texttt{Plain State}: We write the value $0$ to indicate the current value in main state table.
          \item  \texttt{Not Yet Written}: We write the value $1$ to indicate that the slot was never written and its value is zero.
          \item  \texttt{Changeset}: We write the value $2$ to indicate that this value must be read from historical change sets.
        \end{itemize}
        This source annotation allows validators to route prefetch requests optimally. Without source hints, in order to read a storage value at a historical block, the validator must perform the full two-step lookup for every key:
        \begin{enumerate}
          \item History index seek. Seek in Storage History for the (address, slot) pair. This involves a B-tree traversal on a $60$-byte composite key, followed by roaring bitmap decompression and a rank/select operation.
          \item Value fetch. Based on the history index result, read from either Plain Storage State, Storage Change Sets, or return zero.
        \end{enumerate}
        Step 1 is expensive since it is a random B-tree seek into a large index table, and it must be done for every key. We observed that the answer is often "read from Plain State" or "the value is zero." Using Plain State, the validator skips the history index entirely and reads directly from Plain Storage State. This eliminates one B-tree seek per key. Using Not Yet Written, the validator skips all disk I/O. The value is known to be zero without touching the database. Using Changeset, we get no savings, and the validator must perform the full historical lookup.

        As we observe in our analysis, the majority of storage accesses in a typical block hit keys that have not been modified recently, i.e., their current value in Plain State is correct for the historical block being queried. For these keys, the history index lookup is pure overhead: it exists only to confirm that Plain State is the right source. The source hint eliminates that confirmation step. In the best case (Plain State), the hint converts a two-seek operation into a one-seek operation. In the zero case (Not Yet Written), it eliminates disk access entirely. Only keys in the Changeset category still pay full cost.
  \item \italhead{Hint Constructor} Assembles the final Block Hints structure containing storage keys (with sources), account addresses, and bytecode addresses.                                                                                        
  \item \italhead{Hint Writer} Persists hints to a redb database~\cite{bernerCbernerRedb2026} with zstd~\cite{facebookZstandardRealtimeData} compression.
\end{enumerate}

\parhead{Data Flow}
\begin{enumerate}
  \item Block processor fetches block from storage.                                                           
  \item Executor runs transactions with Key Collector inspector.                                               
  \item Key Collector captures all accessed keys via opcode interception.                                      
  \item Source tracker queries storage history to annotate each key with its optimal read source.
  \item Hint constructor assembles the complete hint set.
  \item Hint writer persists hints; state hash writer records execution fingerprint.   
\end{enumerate}

\begin{figure}
  \centering
  \includegraphics[width=\linewidth]{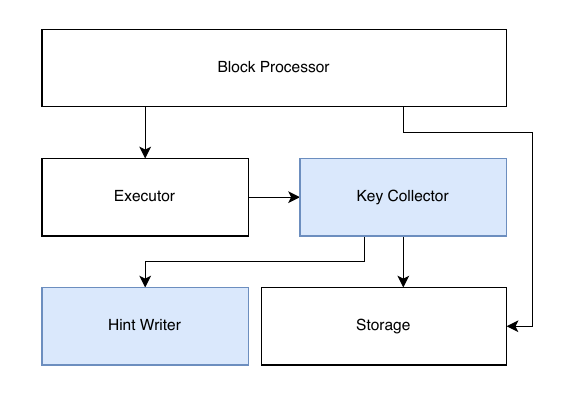}
  \caption{Architecture of our implementation of \sysname{}-L primary. The components marked in blue are the proposed changes to the primary.}
  \label{fig:iral primary architecture}
\end{figure}

\subsection{Reth Backup}%

\emph{reth-backup} is the hint consumer that simulates a validator receiving hints from the proposer. It prefetches all required state before execution, so that block processing incurs zero disk I/O. \Cref{fig:iral backup architecture} presents an illustration of our architecture.

\begin{figure}[!ht]
  \centering
  \includegraphics[width=0.8\linewidth]{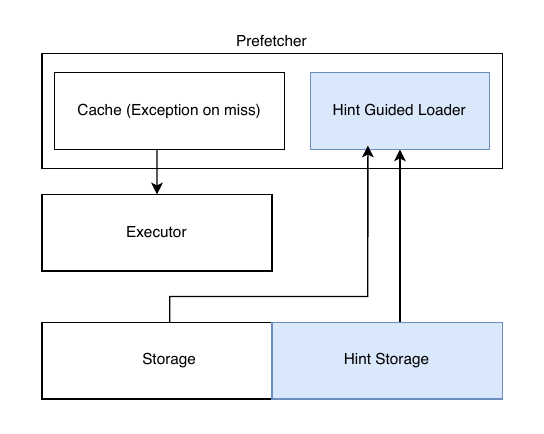}
  \caption{Architecture of our implementation of \sysname{}-L backup. The components marked in blue are proposed changes to the backup.}\label{fig:iral backup architecture}
\end{figure}

\parhead{Components}
\begin{enumerate}[nosep]
  \item \italhead{Prefetcher} Runs on a dedicated thread and loads state ahead of execution in two phases:
        \begin{itemize}[nosep]
          \item \italhead{Phase 1 (startup)}: Prefetches $~3,000$ blocks in parallel using rayon (configurable number of threads), filling an $8$ GB in-memory buffer. This covers the initial cold start.
          \item \italhead{Phase 2 (steady-state)}: Prefetches in sorted batches of $32$ blocks. For each batch, it collects all keys from all blocks, sorts them by address, and reads them using forward cursor walks. This converts random I/O into sequential I/O across the B-tree. Prefetched blocks are sent to the executor via a bounded channel.
        \end{itemize}
        \italhead{Hint-Guided Loader} The core of the prefetcher. For each batch of hints, it routes reads by source annotation:
        \begin{itemize}[nosep]
          \item Plain State: Read directly from the plain state table using a sorted cursor. This is the fast path (~90\% of keys).
          \item Not Yet Written: Insert zero immediately. No disk access.
          \item Changeset: Skipped during prefetch. These rare keys fall back to on-demand reads. 
        \end{itemize}
        The loader also reads account metadata and bytecode, sorting keys before reading to maintain sequential access patterns.

  \item  \italhead{Executor} Consumes pre-populated blocks from the prefetcher. Each block arrives with a cache containing all required state. Execution runs through the EVM with no disk access. The cache is backed by a DB that causes an exception if it does not contain a key: if execution attempts to read a key not in the cache, it crashes, verifying that the hints provide complete coverage. Each cache is discarded after execution to prevent memory accumulation simulating a conservative approach.
        \item \italhead{Storage} The reth database is accessed only by the prefetcher thread, never during execution. Only three tables are used: Plain Account State, Plain Storage State, and Bytecodes.
\end{enumerate}

\parhead{Threading Model}
The prefetcher and executor run on separate threads, connected by a bounded channel (capacity $100$ blocks). This pipelining ensures the executor never waits for I/O: while it executes block $N$, the prefetcher is loading state for block $N+1$ and beyond. The $8$ GB buffer absorbs variance in prefetch and execution times.

\parhead{Key Design Decisions}
\begin{enumerate}
  \item Sorted batch reads. By collecting keys from $32$ blocks, sorting, and reading in order, the prefetcher converts random B-tree seeks into forward cursor walks. This is significantly faster on both SSDs and HDDs.
  \item No cross-block cache. Unlike reth-baseline, the backup discards each block's cache after execution. Hints make the cross-block cache unnecessary since every block's state is prefetched independently.
\end{enumerate}

\fi

\iffullversion
\section{Additional Details}
\label{sec:appendix:details}

\Cref{tab:intra-block} provides detailed bucketed statistics for the intra-block access frequency distribution shown in \Cref{fig:intra-block-frequency}. \Cref{tab:concentration} provides the detailed breakdown for the contract concentration shown in \Cref{fig:contract-concentration}. \Cref{tab:key-lifespan} provides the detailed breakdown for the key lifespan distribution shown in \Cref{fig:key-lifespan}. \Cref{tab:per-block} provides per-block statistics shown in \Cref{fig:per-block-stats}. \Cref{sec:appendix:source-dist} provides the source annotation distribution referenced in \Cref{sec:hint-generation}. \Cref{sec:appendix:backup-speedup} provides the tabulated backup speedup and tail behavior data. \Cref{sec:appendix:hint-cost} provides a detailed analysis of hint cost scaling and variance.

\begin{table}[h]
\centering
\caption{Intra-block access distribution: how many times each key is accessed within its block.}
\label{tab:intra-block}
\small
\begin{tabular}{lrr}
\toprule
\textbf{Accesses per Key} & \textbf{Block-Key Pairs} & \textbf{Share} \\
\midrule
$1$ & $\IntraBlockKeysOne{}$ & $\IntraBlockKeysOnePct{}\%$ \\
$2$ & $\IntraBlockKeysTwo{}$ & $\IntraBlockKeysTwoPct{}\%$ \\
$3$--$5$ & $\IntraBlockKeysThreeToFive{}$ & $\IntraBlockKeysThreeToFivePct{}\%$ \\
$6$--$10$ & $\IntraBlockKeysSixToTen{}$ & $\IntraBlockKeysSixToTenPct{}\%$ \\
$>10$ & $\IntraBlockKeysGtTen{}$ & $\IntraBlockKeysGtTenPct{}\%$ \\
\bottomrule
\end{tabular}
\end{table}

\begin{table}[h]
\centering
\caption{Cumulative storage access share by contract rank.}
\label{tab:concentration}
\small
\begin{tabular}{lr}
\toprule
\textbf{Top-N Contracts} & \textbf{Cumulative Share} \\
\midrule
Top 1 (USDT) & $\ConcentrationTopOne{}\%$ \\
Top 2 (+USDC) & $\ConcentrationTopTwo{}\%$ \\
Top 3 (+Compound) & $\ConcentrationTopThree{}\%$ \\
Top 10 & $\ConcentrationTopTen{}\%$ \\
Top 20 & $\ConcentrationTopTwenty{}\%$ \\
\bottomrule
\end{tabular}
\end{table}

\begin{table}[h]
\centering
\caption{Key lifespan distribution: number of blocks in which each key appears.}
\label{tab:key-lifespan}
\small
\begin{tabular}{lrr}
\toprule
\textbf{Blocks Appeared} & \textbf{Keys} & \textbf{Share} \\
\midrule
$1$ block & $\LifespanKeysOneBlock{}$ & $\LifespanPctOneBlock{}\%$ \\
$2$--$5$ blocks & $\LifespanKeysTwoToFive{}$ & $\LifespanPctTwoToFive{}\%$ \\
$6$--$20$ blocks & $\LifespanKeysSixToTwenty{}$ & $\LifespanPctSixToTwenty{}\%$ \\
$21$--$50$ blocks & $\LifespanKeysTwentyOneToFifty{}$ & $\LifespanPctTwentyOneToFifty{}\%$ \\
$>50$ blocks & $\LifespanKeysGtFifty{}$ & $\LifespanPctGtFifty{}\%$ \\
\bottomrule
\end{tabular}
\end{table}

\begin{table}[h]
\centering
\caption{Per-block storage statistics.}
\label{tab:per-block}
\small
\begin{tabular}{lrrr}
\toprule
\textbf{Metric} & \textbf{Median} & \textbf{P95} & \textbf{Max} \\
\midrule
Total operations & $\TotalOpsMedian{}$ & $\TotalOpsPNinetyFive{}$ & $\TotalOpsMax{}$ \\
Storage operations & $\StorageOpsMedian{}$ & $\StorageOpsPNinetyFive{}$ & $\StorageOpsMax{}$ \\
Unique storage keys & $\UniqueKeysMedian{}$ & $\UniqueKeysPNinetyFive{}$ & $\UniqueKeysMax{}$ \\
\bottomrule
\end{tabular}
\end{table}

\subsection{Per-Block Operation Distribution}
\label{sec:appendix:op-distribution}

\Cref{fig:op-distribution} shows the per-block breakdown of state operations, complementing \Cref{tab:op-distribution}. Storage operations consistently dominate across all blocks.

\begin{figure}[h]
  \centering
  \includegraphics[width=\columnwidth]{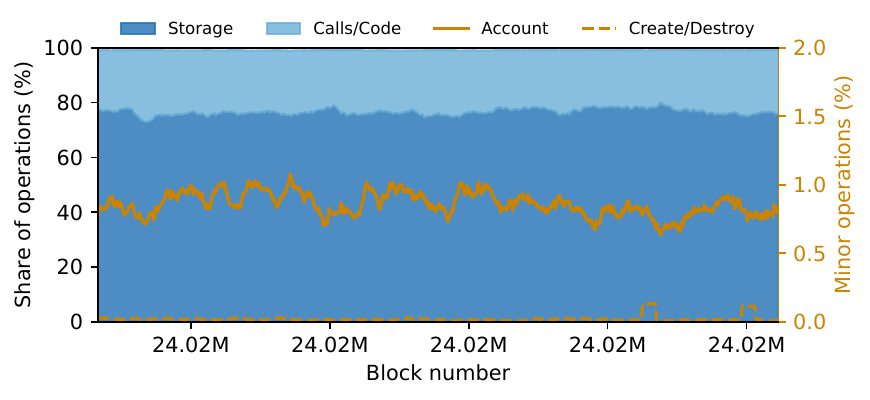}
  \caption{Per-block operation distribution showing storage dominance. Storage operations (blue) consistently account for the majority of state accesses across all blocks.}
  \label{fig:op-distribution}
\end{figure}

\subsection{Source Annotation Distribution}
\label{sec:appendix:source-dist}

Each storage key in the hint carries a one-byte source annotation indicating where the backup should read its value (\Cref{sec:hint-generation}). \Cref{tab:source-dist} reports the aggregate distribution across all $\SourceTotalKeys{}$ storage key accesses in the dataset.

PlainState keys ($\SourcePlainStatePct{}\%$) require a single B-tree seek; Zero keys ($\SourceZeroPct{}\%$) require no I/O at all; only Changeset keys ($\SourceChangesetPct{}\%$) require the full two-step history lookup. In total, $\SourceNoHistoryPct{}\%$ of storage keys bypass the history index entirely.

The per-block distribution is stable: PlainState has a median share of $\SourcePlainStateMedianPct{}\%$ (IQR $\SourcePlainStatePTwentyFivePct{}$--$\SourcePlainStatePSeventyFivePct{}\%$), Zero has a median of $\SourceZeroMedianPct{}\%$ (IQR $\SourceZeroPTwentyFivePct{}$--$\SourceZeroPSeventyFivePct{}\%$), and Changeset has a median of $\SourceChangesetMedianPct{}\%$ (IQR $\SourceChangesetPTwentyFivePct{}$--$\SourceChangesetPSeventyFivePct{}\%$).

\begin{table}[h]
\centering
\caption{Source annotation distribution across storage keys.}
\label{tab:source-dist}
\small
\begin{tabular}{lrrr}
\toprule
\textbf{Source} & \textbf{Keys} & \textbf{Share} & \textbf{Per-Block Median} \\
\midrule
PlainState & $\SourcePlainStateCount{}$ & $\SourcePlainStatePct{}\%$ & $\SourcePlainStateMedianPct{}\%$ \\
Zero & $\SourceZeroCount{}$ & $\SourceZeroPct{}\%$ & $\SourceZeroMedianPct{}\%$ \\
Changeset & $\SourceChangesetCount{}$ & $\SourceChangesetPct{}\%$ & $\SourceChangesetMedianPct{}\%$ \\
\midrule
No history (PlainState + Zero) & $-$ & $\SourceNoHistoryPct{}\%$ & $\SourceNoHistoryMedianPct{}\%$ \\
\bottomrule
\end{tabular}
\end{table}

\subsection{Backup Speedup Details}
\label{sec:appendix:backup-speedup}

\Cref{tab:speedup-summary} provides the per-block speedup percentiles summarizing \Cref{fig:speedup-per-block}. \Cref{tab:regression} provides the tail behavior breakdown referenced in \Cref{sec:eval:tail}.

\begin{table}[h]
\centering
\caption{Per-block speedup distribution (backup vs.\ baseline).}
\label{tab:speedup-summary}
\small
\begin{tabular}{lrrrr}
\toprule
\textbf{Configuration} & \textbf{Median} & \textbf{P90} & \textbf{P99} & \textbf{Max} \\
\midrule
Sequential   & $\EvalSpeedupSequentialMedian{}\times$ & $\EvalSpeedupSequentialPNinety{}\times$ & $\EvalSpeedupSequentialPNinetyNine{}\times$ & $\EvalSpeedupSequentialMax{}\times$ \\
Parallel-16  & $\EvalSpeedupParallelSixteenMedian{}\times$ & $\EvalSpeedupParallelSixteenPNinety{}\times$ & $\EvalSpeedupParallelSixteenPNinetyNine{}\times$ & $\EvalSpeedupParallelSixteenMax{}\times$ \\
Parallel-64  & $\EvalSpeedupParallelSixtyFourMedian{}\times$ & $\EvalSpeedupParallelSixtyFourPNinety{}\times$ & $\EvalSpeedupParallelSixtyFourPNinetyNine{}\times$ & $\EvalSpeedupParallelSixtyFourMax{}\times$ \\
\bottomrule
\end{tabular}
\end{table}

\begin{table}[h]
\centering
\caption{Fraction of blocks where backup is slower than baseline.}
\label{tab:regression}
\small
\begin{tabular}{lrr}
\toprule
\textbf{Configuration} & \textbf{Slower Blocks} & \textbf{Fraction} \\
\midrule
Sequential   & $\EvalSlowerSequentialCount{}$ & $\EvalSlowerSequentialPct{}\%$ \\
Parallel-16  & $\EvalSlowerParallelSixteenCount{}$    & $\EvalSlowerParallelSixteenPct{}\%$ \\
Parallel-64  & $\EvalSlowerParallelSixtyFourCount{}$     & $\EvalSlowerParallelSixtyFourPct{}\%$ \\
\bottomrule
\end{tabular}
\end{table}

\subsection{Hint Cost Variance Analysis}
\label{sec:appendix:hint-cost}

The per-block hint fraction (\Cref{fig:per-block-hint-cost}) varies across blocks despite a stable per-key cost. \Cref{fig:per-block-hint-cost-absolute} plots the same distributions in absolute time: construction spans a wide range while serialization clusters tightly around ${\sim}\EvalSerializationMedianMs{}$\,ms. \Cref{fig:hint-cost-scaling} confirms that construction scales linearly with working set size ($r = \EvalConstructKeysCorr{}$, ${\sim}\EvalConstructPerKeyUs{}$\,$\mu$s/key), while serialization is nearly constant ($r = \EvalWriteKeysCorr{}$).

\begin{figure}[h]
  \centering
  \includegraphics[width=\columnwidth]{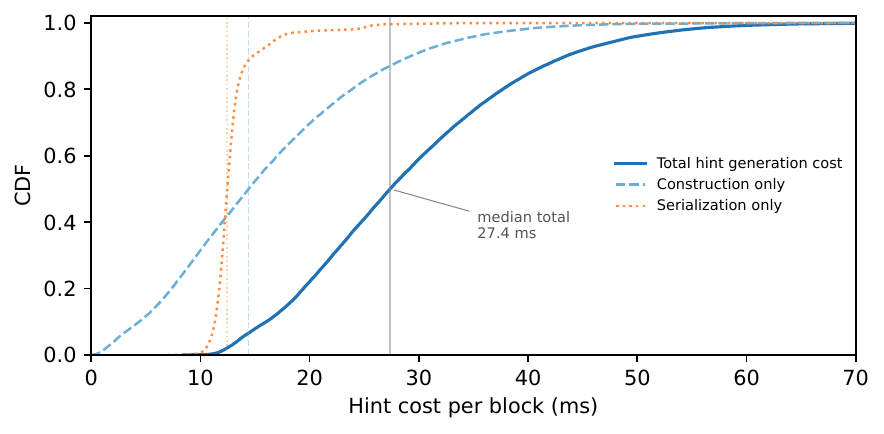}
  \caption{CDF of per-block hint cost in absolute time (ms). Construction time varies widely with block size, while serialization clusters tightly around ${\sim}\EvalSerializationMedianMs{}$\,ms.}
  \label{fig:per-block-hint-cost-absolute}
\end{figure}

\begin{figure}[h]
  \centering
  \includegraphics[width=\columnwidth]{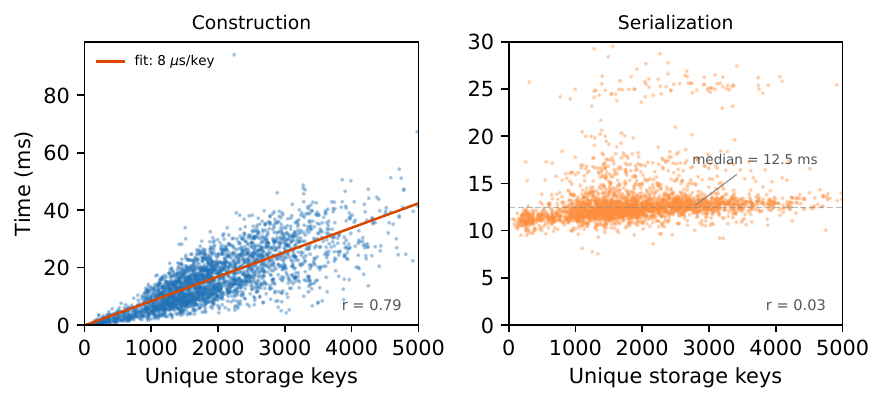}
  \caption{Hint cost scaling with block working set size. Construction time (left) scales linearly with the number of unique storage keys ($r = \EvalConstructKeysCorr{}$, ${\sim}\EvalConstructPerKeyUs{}$\,$\mu$s/key). Serialization time (right) is nearly constant regardless of block size ($r = \EvalWriteKeysCorr{}$, median $\EvalSerializationMedianMs{}$\,ms).}
  \label{fig:hint-cost-scaling}
\end{figure}

\Cref{tab:hint-cost-bins} shows how the hint fraction changes with block working set size. The median hint fraction decreases monotonically from $\EvalBinLtFiveHundredHintFracMedian{}\%$ for small blocks ($<$500 keys) to $\EvalBinGtFiveKHintFracMedian{}\%$ for large blocks ($>$5K keys), even though absolute hint time increases. This occurs because execution time grows faster than hint time for larger blocks: large blocks perform more I/O, and the I/O-dominated execution time scales super-linearly with working set size while hint construction scales linearly (\Cref{fig:hint-cost-scaling}).

\begin{table}[h]
\centering
\caption{Hint cost breakdown by block working set size.}
\label{tab:hint-cost-bins}
\small
\begin{tabular}{lrrrr}
\toprule
\textbf{Unique Keys} & \textbf{Blocks} & \textbf{Hint Frac.} & \textbf{Exec Time} & \textbf{Hint Time} \\
 & & \textbf{(median)} & \textbf{(median)} & \textbf{(median)} \\
\midrule
$<$500 & $\EvalBinLtFiveHundredCount{}$ & $\EvalBinLtFiveHundredHintFracMedian{}\%$ & $\EvalBinLtFiveHundredExecMedianMs{}$ ms & $\EvalBinLtFiveHundredHintMedianMs{}$ ms \\
500--1K & $\EvalBinFiveHundredToOneKCount{}$ & $\EvalBinFiveHundredToOneKHintFracMedian{}\%$ & $\EvalBinFiveHundredToOneKExecMedianMs{}$ ms & $\EvalBinFiveHundredToOneKHintMedianMs{}$ ms \\
1K--2K & $\EvalBinOneKToTwoKCount{}$ & $\EvalBinOneKToTwoKHintFracMedian{}\%$ & $\EvalBinOneKToTwoKExecMedianMs{}$ ms & $\EvalBinOneKToTwoKHintMedianMs{}$ ms \\
2K--3K & $\EvalBinTwoKToThreeKCount{}$ & $\EvalBinTwoKToThreeKHintFracMedian{}\%$ & $\EvalBinTwoKToThreeKExecMedianMs{}$ ms & $\EvalBinTwoKToThreeKHintMedianMs{}$ ms \\
3K--5K & $\EvalBinThreeKToFiveKCount{}$ & $\EvalBinThreeKToFiveKHintFracMedian{}\%$ & $\EvalBinThreeKToFiveKExecMedianMs{}$ ms & $\EvalBinThreeKToFiveKHintMedianMs{}$ ms \\
$>$5K & $\EvalBinGtFiveKCount{}$ & $\EvalBinGtFiveKHintFracMedian{}\%$ & $\EvalBinGtFiveKExecMedianMs{}$ ms & $\EvalBinGtFiveKHintMedianMs{}$ ms \\
\bottomrule
\end{tabular}
\end{table}

\parhead{Variance decomposition}
To identify the dominant source of per-block hint fraction variance, we fix each component at its median and measure the resulting variance reduction. Fixing execution time at its median reduces hint fraction variance by $\EvalHintFracVarFromExecPct{}\%$, confirming that I/O-driven execution time variability, not hint cost variability, is the primary source of per-block fluctuations. Fixing serialization time at its median reduces variance by $\EvalHintFracVarFromWritePct{}\%$, consistent with the observation that serialization exhibits occasional latency spikes from disk contention while maintaining a tight steady-state distribution (IQR $\EvalSerializationIQRLowMs{}$-$\EvalSerializationIQRHighMs{}$ ms).

\parhead{Serialization cost}
Serialization time is dominated by the fixed overhead of committing the database write transaction (opening, flushing, and closing the redb write batch), not by the volume of data written. This explains the near-zero correlation with block size ($r = \EvalWriteKeysCorr{}$). The tight IQR ($\EvalSerializationIQRLowMs{}$--$\EvalSerializationIQRHighMs{}$ ms) indicates stable steady-state performance, with rare spikes attributable to background disk I/O contention from the concurrent MDBX database.
\fi

\section{Extended Related Work}
\label{sec:appendix:extended-related}
\parhead{Blockchain State Synchronization (Extended)}
Early Ethereum clients performed full sync by re-executing every transaction from genesis, but this approach became impractical as chain history grew. Geth~\cite{EthereumGoethereum2026} introduced fast sync~\cite{Eth63Fast} (2015) and later snap sync~\cite{GethV1100} (2021), which download recent state directly and reconstruct the Merkle trie locally, achieving $10\times$ faster initial sync. Erigon~\cite{ErigontechErigon2026} (formerly Turbo-Geth) pioneered \emph{staged sync}~\cite{ErigonEthStagedsync}, a pipelined architecture that processes headers, bodies, senders recovery, and execution as discrete stages; Reth~\cite{ParadigmxyzReth2026} adopts this architecture in Rust for further performance gains. Despite these optimizations, the execution stage cannot be parallelized within staged sync.

Regarding access lists, analysis shows only $1.46\%$ of transactions include access lists despite $42.6\%$ benefiting from them~\cite{heimbachDissectingEIP2930Optional2023}. Solana's Sealevel runtime~\cite{yakovenkoSolanaNewArchitecture2025} mandates read/write set declarations, enabling lock-based parallel execution across non-conflicting transactions. Block-STM~\cite{gelashviliBlockSTMScalingBlockchain2022} achieves 170k transactions per second on Aptos using software transactional memory techniques. Sui~\cite{blackshearSuiLutrisBlockchain2024} takes an object-centric approach where owned objects bypass consensus entirely, enabling sub-second finality for non-shared state. Forerunner~\cite{chenForerunnerConstraintbasedSpeculative2021} exploits the time window between transaction dissemination and consensus to speculatively pre-execute transactions, achieving $10\times$ speedup through constraint-based multi-trace specialization. Looking forward, Verkle trees~\cite{kuszmaulVerkleTrees2018} promise to enable stateless clients by reducing witness sizes from ${\sim}150$ KB to $1$--$2$ KB, allowing validators to verify blocks without storing full state.

\parhead{Speculative and Parallel Execution (Extended)}
For workloads with high contention on popular records, Doppel~\cite{narulaPhaseReconciliationContended2014} introduced phase reconciliation, cycling between split phases (where commutative operations execute on per-core state) and joined phases (where results are merged), achieving up to $38\times$ higher throughput than conventional protocols. Deterministic approaches like BOHM~\cite{faleiroRethinkingSerializableMultiversion2015} pre-compute transaction dependencies to enable parallel execution without runtime conflict detection, while PWV~\cite{faleiroHighPerformanceTransactions2017} increases concurrency by making writes visible before transaction commit through piece-wise execution. In the blockchain space, commercial efforts such as Monad~\cite{IntroductionMonadDeveloper} and Sei~\cite{SeiParallelizationEngine2026} further pursue optimistic parallel EVM execution. These works share a common goal: maximizing transaction throughput by exploiting parallelism. \sysname{} is orthogonal to and composable with all of these approaches.

\iffullversion
\section{Generalized \sysname{} Architecture}
\label{sec:appendix:general-ira}

In this appendix, we present the generalized \sysname{} protocol that can be adapted to any primary-backup system with a key-value store interface, and provide sketches for integrating \sysname{} into production systems.

\subsection{Abstract Protocol}

\parhead{System Model}
Consider a primary-backup replication system where:
\begin{itemize}[nosep]
  \item State is stored as a key-value map $S: \mathcal{K} \to \mathcal{V}$
  \item Transactions are batches of operations $T = \langle op_1, \ldots, op_n \rangle$ where each $op_i$ is either $\textsc{Read}(k)$ or $\textsc{Write}(k, v)$
  \item The primary executes transactions and propagates them to backups via a replicated log
  \item Backups replay transactions from the log to maintain consistent state
\end{itemize}

\parhead{Hint Generation (Primary)}
During execution of transaction batch $T$, the primary maintains an access set $A = \emptyset$. For each operation:
\begin{itemize}[nosep]
  \item $\textsc{Read}(k)$: Add $k$ to $A$
  \item $\textsc{Write}(k, v)$: Add $k$ to $A$ (writes typically require reading the old value for logging or conflict detection)
\end{itemize}
After execution, the primary constructs hint $H = (A, M)$ where $M$ is optional metadata (source annotations, access ordering, etc.). The primary transmits $H$ alongside or ahead of $T$ to backups.

\parhead{Hint-Based Prefetch (Backup)}
Upon receiving hint $H = (A, M)$, the backup:
\begin{enumerate}
  \item Sorts $A$ by key order to enable sequential I/O through the storage engine's B-tree or LSM index
  \item For each $k \in A$, fetches $S[k]$ into an in-memory cache
  \item If $M$ contains source annotations, routes reads to the appropriate storage tier (e.g., main table vs.\ historical snapshots)
\end{enumerate}

\parhead{Replay (Backup)}
The backup executes transaction $T$ with all keys pre-cached. All $\textsc{Read}(k)$ operations return immediately from cache; $\textsc{Write}(k, v)$ operations update the cache and mark entries dirty for eventual persistence.

\Cref{alg:ira-general-primary,alg:ira-general-backup} present the generalized hint generation and replay algorithms.

\begin{algorithm}[t]
\caption{Generalized \sysname{} Hint Generation (Primary)}
\label{alg:ira-general-primary}
\begin{algorithmic}[1]
\Require Transaction batch $T = \langle op_1, \ldots, op_n \rangle$, State $S$
\Ensure Hint $H$, Execution result $R$
\Statex
\State $A \gets \emptyset$ \Comment{Access set}
\For{$i \gets 1$ \textbf{to} $n$}
    \If{$op_i = \textsc{Read}(k)$}
        \State $A \gets A \cup \{k\}$
        \State Execute read from $S$
    \ElsIf{$op_i = \textsc{Write}(k, v)$}
        \State $A \gets A \cup \{k\}$
        \State Execute write to $S$
    \EndIf
\EndFor
\Statex
\State $M \gets \textsc{ComputeMetadata}(A, S)$ \Comment{Optional: sources, ordering}
\State $H \gets (A, M)$
\State \Return $(H, R)$
\end{algorithmic}
\end{algorithm}

\begin{algorithm}[t]
\caption{Generalized \sysname{} Replay (Backup)}
\label{alg:ira-general-backup}
\begin{algorithmic}[1]
\Require Transaction batch $T$, Hint $H = (A, M)$, State $S$
\Ensure Updated state $S'$
\Statex
\LineComment{Phase 1: Prefetch via sorted sequential scan}
\State $A' \gets \textsc{SortByKey}(A)$
\State $\mathit{cache} \gets \emptyset$
\For{$k \in A'$}
    \State $\mathit{cache}[k] \gets S[k]$ \Comment{Sequential I/O}
\EndFor
\Statex
\LineComment{Phase 2: Execute with all state in cache}
\For{$op \in T$}
    \If{$op = \textsc{Read}(k)$}
        \State \Return $\mathit{cache}[k]$ \Comment{Cache hit}
    \ElsIf{$op = \textsc{Write}(k, v)$}
        \State $\mathit{cache}[k] \gets v$
        \State Mark $k$ as dirty
    \EndIf
\EndFor
\Statex
\LineComment{Phase 3: Persist dirty entries}
\For{$k \in \mathit{cache}.\textsc{Dirty}()$}
    \State $S[k] \gets \mathit{cache}[k]$
\EndFor
\end{algorithmic}
\end{algorithm}

\subsection{Hint Transmission Strategies}

The protocol admits several transmission strategies with different tradeoffs:

\parhead{Inline Transmission}
Transmit hint $H$ as part of the log entry containing transaction batch $T$. This is the simplest approach: hints and transactions are atomically committed and delivered together. The cost is increased log entry size, which may affect replication latency for bandwidth-constrained networks.

\parhead{Sideband Transmission}
Transmit hint $H$ through a separate channel ahead of transaction $T$. This allows backups to begin prefetching before the transaction arrives, hiding prefetch latency behind network transmission time. The cost is coordination complexity: backups must match hints to their corresponding transactions, and hints may arrive out of order or be lost independently.

\parhead{On-Demand Transmission}
Backups request hints from the primary only when they detect high cache miss rates. This reduces bandwidth when the workload exhibits high locality (most keys are already cached), but increases latency when hints are needed. Suitable for systems with variable workload patterns.

\subsection{System-Specific Integration Sketches}

\parhead{Apache Kafka (Kafka Streams)}
Kafka Streams applications maintain local state stores (typically backed by RocksDB) that are updated during record processing. When a consumer falls behind, it must replay records while performing state store lookups that may miss the RocksDB block cache.

\italhead{Integration}
The producing application, which knows what state store keys each record will access, attaches the access set as a record header. Consumers extract the header and prefetch state store entries before processing each batch. For changelog-backed state stores, hints can be derived from the changelog records themselves, with each changelog entry implying a future read of that key.

\parhead{Raft-Based Systems (etcd, TiKV, CockroachDB)}
In Raft, the leader executes commands and replicates log entries to followers. Followers apply log entries to their local state machine, typically backed by RocksDB or a similar embedded database.

\italhead{Integration}
The leader's state machine execution already observes key accesses. Extend the Raft log entry format to include an optional access set field. Followers extract the access set and prefetch from their local RocksDB instance before applying the entry. This is particularly valuable during leader failover: the new leader must replay any uncommitted entries, and hint-based prefetching accelerates this recovery path.

\parhead{PostgreSQL Streaming Replication}
PostgreSQL standbys apply Write-Ahead Log (WAL) records streamed from the primary. Each WAL record describes changes to specific heap pages and index pages. Replay performance depends on whether the required pages are in the standby's buffer pool.

\italhead{Integration}
The primary's buffer manager observes which pages are accessed during transaction execution. Extend the WAL protocol to include page-level access lists, either per-record or aggregated per checkpoint interval. Standbys use these lists to prefetch pages from disk into their buffer pool before applying WAL records. This is especially valuable for standbys with smaller buffer pools than the primary or after standby restart when the buffer pool is cold.

\parhead{MySQL Group Replication / Galera Cluster}
MySQL replication uses binary log (binlog) events that describe row-level changes. Replicas apply these events by locating affected rows through index lookups, which may incur cache misses in the InnoDB buffer pool.

\italhead{Integration}
During binlog event creation, the primary observes which index pages are traversed for each row operation. Extend the replication protocol to include these page identifiers as hints. Replicas prefetch the indicated pages into their buffer pool before applying the binlog event. For certification-based replication (Galera), hints can be attached to the writeset certification request.

\parhead{Distributed Key-Value Stores (FoundationDB, TiKV)}
These systems use consensus-based replication where storage servers replay the leader's mutation log. The leader (or transaction coordinator) knows exactly which keys each transaction reads and writes.

\italhead{Integration}
FoundationDB's storage servers receive mutation batches from the transaction log. Extend the mutation batch format to include the read set (keys read during transaction execution) in addition to the write set. Storage servers prefetch read-set keys from their local SQLite or RocksDB instance before applying mutations. TiKV can similarly extend its Raft log entries with read-set hints, enabling followers to prefetch before applying proposals.

\subsection{Hint Sizing and Bandwidth Considerations}

\parhead{Key Representation}
For systems with structured keys (e.g., $(table, row, column)$ tuples or composite keys with common prefixes), hints can be compressed:
\begin{itemize}[nosep]
  \item \textbf{Prefix compression:} Many keys share common prefixes (same table, adjacent rows). Store the common prefix once and encode suffixes differentially.
  \item \textbf{Bloom filters:} For approximate prefetching, encode the access set as a Bloom filter. False positives cause unnecessary prefetches (wasted I/O) but false negatives are impossible and all the accessed keys will be prefetched.
  \item \textbf{Range encoding:} For range scans, encode $(start, end)$ rather than enumerating individual keys.
\end{itemize}

\parhead{Bandwidth-Latency Tradeoff}
Let $B_H$ be the hint bandwidth cost (bytes per transaction) and $\Delta L$ be the replay latency reduction from hint-based prefetching. The system benefits from hints when:
\[
\frac{B_H}{\mathit{bandwidth}} < \Delta L \cdot |\mathit{backups}|
\]
For high-fanout replication (many backups), even moderately expensive hints become worthwhile because the primary generates hints once but multiple backups benefit. \sysname{}-L (\Cref{sec:ira-l}) achieves median hint sizes of $\EvalHintCompressedMedianKB{}$ KB per block which is comparable to a few TCP packets, while providing $\EvalSpeedupSequentialMedian{}\times$ median speedup.

\parhead{Adaptive Hint Granularity}
Systems can dynamically adjust hint detail based on observed conditions:
\begin{itemize}[nosep]
  \item \textbf{High cache hit rate:} Reduce hint detail or disable hints entirely. The workload has sufficient locality that standard caching policies perform well.
  \item \textbf{High cache miss rate:} Increase hint detail (add source annotations, access ordering). The workload is random or the working set exceeds cache capacity.
  \item \textbf{Network congestion:} Switch to lossy representations (Bloom filters, sampled access sets) that trade prefetch accuracy for reduced bandwidth.
\end{itemize}

\fi

\end{document}